\documentclass{article}

\usepackage[utf8]{inputenc}

\usepackage{amsfonts,amssymb,amsmath}
\usepackage{graphicx}
\usepackage[english]{babel}
\usepackage{authblk}
\usepackage{enumitem}
\usepackage{numprint}
\usepackage{float}
\usepackage{placeins}
\usepackage[numbers]{natbib}
\usepackage{xr-hyper}
\usepackage{hyperref}
\usepackage[capitalise]{cleveref}
\usepackage{subcaption}
\usepackage{xstring}
\usepackage{setspace} 
\usepackage{microtype}
\usepackage{booktabs}
\usepackage{longtable}
\usepackage{rotating}
\usepackage{fancyhdr}
\usepackage{multirow}
\usepackage{colortbl}
\usepackage{pgfplotstable}
\usepackage{bm}
\usepackage{csquotes}
\usepackage{kky}
\usepackage{xcolor}
\usepackage{textcomp}
\usepackage{phonetic}
\usepackage{ulem}

\newcommand{\corrected}[1]{#1}
\newcommand{\correctednn}[1]{#1}%

\renewcommand{\sout}[1]{}

\DeclareUnicodeCharacter{00B0}{ }
\DeclareCaptionFormat{myformat}{\textbf{#1#2 $|$} #3}
    {\clearpage\pagebreak[4]\global\pdfpageattr\expandafter{\the\pdfpageattr/Rotate 90}}%
    {\clearpage\pagebreak[4]\global\pdfpageattr\expandafter{\the\pdfpageattr/Rotate 0}}%

\newcommand{\papertitle}{Multi-omics Prediction from High-content Cellular Imaging with Deep Learning}
\title{\papertitle}
\author[1,a]{Rahil Mehrizi}
\author[1,a]{Arash Mehrjou}
\author[1]{Maryana Alegro}
\author[1]{Yi Zhao}
\author[1]{Benedetta Carbone}
\author[1]{Carl Fishwick}
\author[1]{Johanna Vappiani}
\author[1]{Jing Bi}
\author[1]{Siobhan Sanford}
\author[1]{Hakan Keles}
\author[1]{Erin Edwards}
\author[1]{Guillaume Heger}
\author[1]{Mathias Kalxdorf}
\author[1]{Marcus Bantscheff}
\author[1]{Cuong Nguyen}
\author[1,*]{Patrick Schwab}
\affil[1]{GSK plc, United Kingdom}
\affil[a]{Joint first authors}
\affil[*]{Corresponding author}

\date{}
\makeatletter
\renewcommand{\@seccntformat}[1]{}
\makeatother

\newcommand{\themethod}{Image2Omics}

\begin{document}

\maketitle
\thispagestyle{fancy}
\pagestyle{fancy}

\vspace{-2.5em}
\section{Abstract}

High-content cellular imaging, transcriptomics, and proteomics data provide rich and complementary views on the molecular layers of biology that influence cellular states and function.  However, the biological determinants through which changes in multi-omics measurements influence cellular morphology have not yet been systematically explored, and the degree to which cell imaging could potentially enable the prediction of multi-omics directly from cell imaging data is therefore currently unclear. Here, we address the question of whether it is possible to predict bulk multi-omics measurements directly from cell images using \themethod{} -- a deep learning approach that predicts multi-omics in a cell population directly from high-content images \corrected{of cells} stained with multiplexed fluorescent dyes. We perform an experimental evaluation in gene-edited macrophages derived from human induced pluripotent stem cells (hiPSC) under multiple stimulation conditions and demonstrate that \themethod{} achieves significantly better performance in predicting transcriptomics and proteomics measurements directly from cell images than \corrected{predictions} based on the mean observed training set abundance. 
We observed significant predictability of abundances for \correctednn{4927 (18.72\%; 95\% CI: 6.52\%, 35.52\%) and 3521 (13.38\%; 95\% CI: 4.10\%, 32.21\%) transcripts out of 26137 in M1 and M2-stimulated macrophages respectively and for 422 (8.46\%; 95\% CI: 0.58\%, 25.83\%) and 697 (13.98\%; 95\% CI: 2.41\%, 32.83\%) proteins out of 4986} in M1 and M2-stimulated macrophages respectively.
Our results show that some transcript and protein abundances are predictable from cell imaging and that cell imaging may potentially, in some settings and depending on the mechanisms of interest and desired performance threshold, even be a scalable and resource-efficient substitute for multi-omics measurements.  \newpage

\section{Introduction}
\label{sec:introduction}

 \begin{figure}
     \centering
     \includegraphics[width=0.90\linewidth]{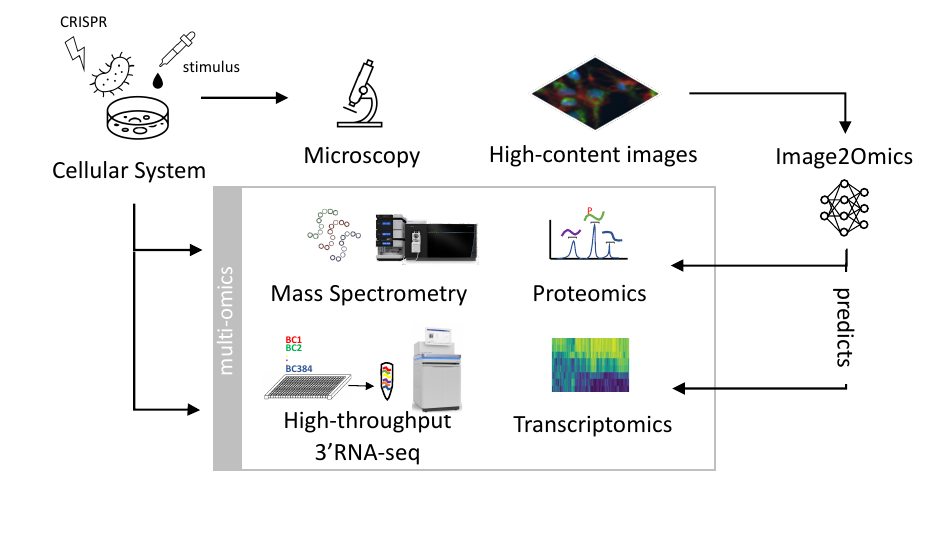}
     \caption{\textbf{Predicting multi-omics from high content cellular imaging.} We generated cell imaging data for a cellular system under a wide range of CRISPR perturbations and exposed to multiple stimuli (top left; in this work: M1- and M2-polarised macrophages). We then trained a machine learning model (\themethod{}; right) using the paired samples where both imaging and multi-omics (transcriptomics and proteomics) were available to learn how to predict the multi-omics layers directly from high-content images alone \correctednn{using independently fine-tuned models for each omics modality and stimulation condition}.}
     \label{fig:fig1}
 \end{figure}

Transcriptomics, proteomics and high-content imaging are rich and complementary tools for interrogating state and function of cellular systems at high-resolution across multiple layers of biology \cite{bray2016cell,hasin2017multi,pinu2019systems,subramanian2020multi}. However, measuring transcriptomics and proteomics -- even with optimised technologies \cite{li2022drugseq,gillet2016mass,zampieri2017frontiers} - is resource- and time-intensive \cite{mantione2014comparing} and often the main bottleneck when conducting studies aiming to elucidate molecular biology. High-content imaging on the other hand is a high-throughput technology that can generate measurements at scale in a cost- and time-effective manner \cite{starkuviene2007potential,bray2017dataset,hua2021cytoimagenet,way2021predicting}. 
While at any given snapshot in time, capturing multiple omics layers (including proteomics, transcriptomics, and imaging) may offer nuanced advantages in that each layer may capture a distinct segment of the complete gene to function pathway, cell supply considerations may limit the feasibility of doing so. As discovery efforts increasingly turn to more complex in vitro models of disease, cell models such as stem cells, primary cells, or even 3D cell culture models (i.e. organoids) may not be reasonably scalable or at least not without extensive investment in automation and other resources. Therefore, the use of the highest-throughput and least expensive of these technologies to faithfully predict the others would represent a significant advancement in the field of drug discovery.

Given that molecular biology and cell morphology\footnote{For brevity, throughout this work, we refer to cell morphology as a term that includes both image-derived morphological as well as protein localisation features from antibody staining.} regulate each other in a complex interplay \cite{wada2011hippo,hamilton2001regulation,canton2006shape,tsuji2018morphology}, we hypothesise that high-content images of cells under controlled conditions contain information that enables the approximate reconstruction of the underlying molecular state. However, to the best of our knowledge, whether and to what degree the various multi-omics layers can be reconstructed from high-content imaging data has not yet been established. The ability to reconstruct molecular biological measurements with multi-omics technologies could aid in optimising experimental throughput while potentially retaining the same information content at lower resource expenditure.

Establishing the predictability of multi-omics from cell images is difficult because (i) both cell images and multi-omics layers are rich in information content and manually interpreting their contents for associations with underlying omics markers is infeasible and prone to over- and mis-interpretation of mechanisms, and (ii) a sufficiently representative set of reference images with omics annotations is necessary to evaluate prediction performance across the diversity of cell states in a target cell population. Morphological features corresponding to high-level function, such as elongated shapes in M1/M2-polarized macrophages \cite{mcwhorter2015physical} and the various characteristic shapes adopted by microglia in healthy and diseased conditions \cite{savage2019morphology}, are established in literature, but - to the best of our knowledge - no comprehensive map relating cellular morphology and protein localisation to the underlying molecular biology exists to date.

The translatability between transcriptomics measurements and high-content imaging has previously been studied in the single-cell context by \cite{yang_multi-domain_2021} - where multi-domain autoencoders are used to translate between single-cell imaging and transcriptomics readouts - and by \cite{Lee2022morphnet}, wherein a model for generating nuclear and whole-cell morphology from single-cell gene expression profiles was developed. Outside of high-content cell imaging, the predictability of transcriptomics was previously demonstrated from hematoxylin and eosin (H\&E) stained histopathology images \cite{Arslan2022multihe,schmauch_deep_2020,azuaje2019connecting}, from positron emission tomography (PET) data \cite{tixier_transcriptomics_2020}, from computerized tomography (CT) data \cite{aonpong_genotype-guided_2021}, from computerized tomography angiography (CTA) data \cite{buckler2021virtual} and from imaging flow cytometry data \cite{chlis_predicting_2020}. In addition, \cite{peng_gluer_2021,bao_integrative_2022} introduced methods for integrating single-cell multi-omics measurements and imaging data \cite{watson_computational_2022,adossa_computational_2021}, and a large-scale dataset of paired gene expression and morphological profiles of cells under perturbations was released by \cite{Haghighi2021dataset}. However, to the best of our knowledge, the predictability of multi-omics measurements directly from high-content cell imaging data has not yet been studied.

In this work, we demonstrate that transcriptomics and proteomics measurements can be predicted directly from high-content imaging data using a machine learning model, \themethod{}, trained on paired multi-omics measurements and cell images collected in hiPSC-derived macrophage\corrected{s} under two stimulation conditions and under 152 CRISPR-based perturbations to cover a diverse range of cellular states. To the best of our knowledge, this work is the first to demonstrate that the prediction of multi-omics from cell imaging can be achieved across transcriptome and proteome and under multiple stimulation conditions. Our results imply that high-content imaging data is predictive of transcriptomics and proteomics measurements, and may - in some settings and depending on the mechanisms of interest -- be an appropriate substitute for multi-omics measurements, particularly if a paired omics and imaging dataset in the same cell population under comparable conditions is available \corrected{for training}.

\section{Results} %
\label{sec:results}
\paragraph{\themethod{}.} \themethod{} is a machine-learning model trained to predict bulk transcriptomics and proteomics directly from high-content images. \themethod{} uses a multiple instance learning approach to train a residual network (ResNet) \cite{he2016deep} architecture deep convolutional neural network with approximately 12 million parameters. Each well image is divided into smaller regions (tiles) that contain a single cell in the center. The ResNet architecture takes 64 randomly selected cell-centered tiles as input and produces a latent representation for each well image by aggregating the results across multiple tiles. The transcript and protein abundance measurements are predicted \correctednn{independently for each omics layer and stimulation condition} from the well latent representations (\Cref{sec:imageanalysis}). To collect the training and test data for developing and evaluating \themethod{}, hiPSC-derived \cite{malik2013review} \corrected{macrophages were perturbed using clustered regularly interspaced short palindromic repeats (CRISPR) technology to knock out 152 genes, and cells were then stimulated towards pro- (M1) or anti- (M2) inflammatory states, thus resulting in} a diverse range of cell states. Measurements were subsequently taken for bulk proteomics, bulk transcriptomics, and high-content imaging (see \Cref{sec:data_acquisition} for details). We empirically evaluated the performance of \themethod{} {in} predicting bulk transcriptomics and proteomics measurements from high-content imaging data by comparing the predictions made by \themethod{} to measured ground truth observations in held-out test set across 10 different random splits (\Cref{sec:evaluation_protocol}).

\begin{figure}
    \centering
    \includegraphics[width=0.81\textwidth]{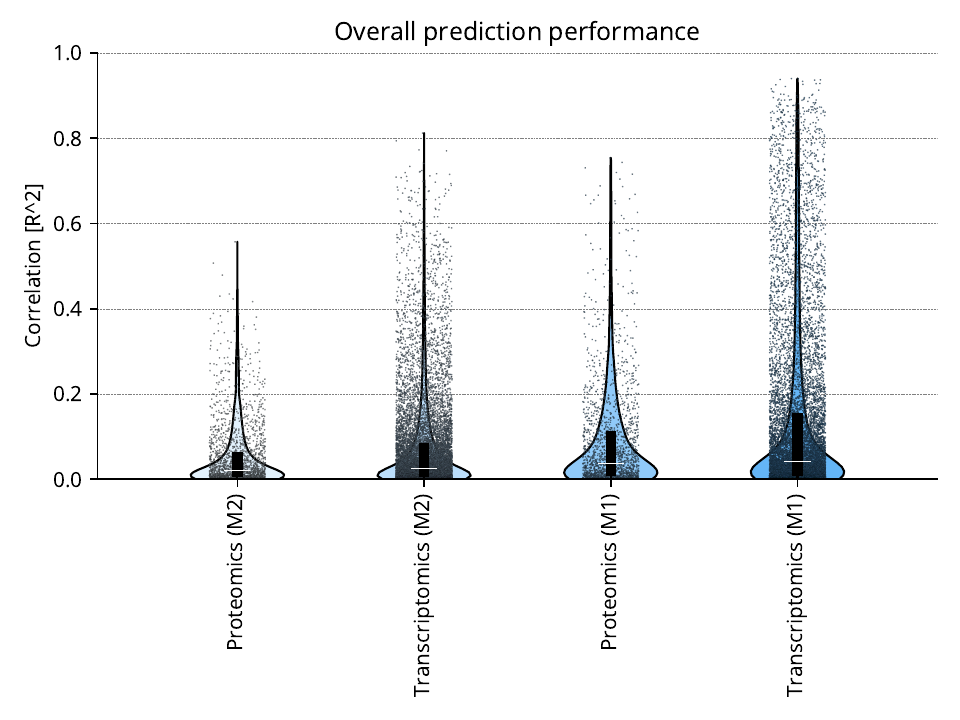}\vspace{1em}\quad
       \caption{\textbf{Overall prediction performance.} Correlation coefficients ($r^2$; y-axis, higher = more predictable) between observed protein and transcript abundances and those predicted by \themethod{} in M1 and M2 polarised states on all held out test set samples. Dots correspond to transcript or protein markers and violins indicate the distribution of coefficient of determinations across the transcriptome and proteome in M1 and M2 states. A number of selected genes from the top and bottom 10 for each stimulus and gene product with the respectively lowest and highest prediction errors are available in \Cref{tb:top_genes_performances_1,tb:top_genes_performances_2,tb:top_genes_performances_3,tb:top_genes_performances_4}. }
       \label{fig:overall_performances}
\end{figure}

\begin{figure}
    \centering
    \includegraphics[width=0.48\textwidth]{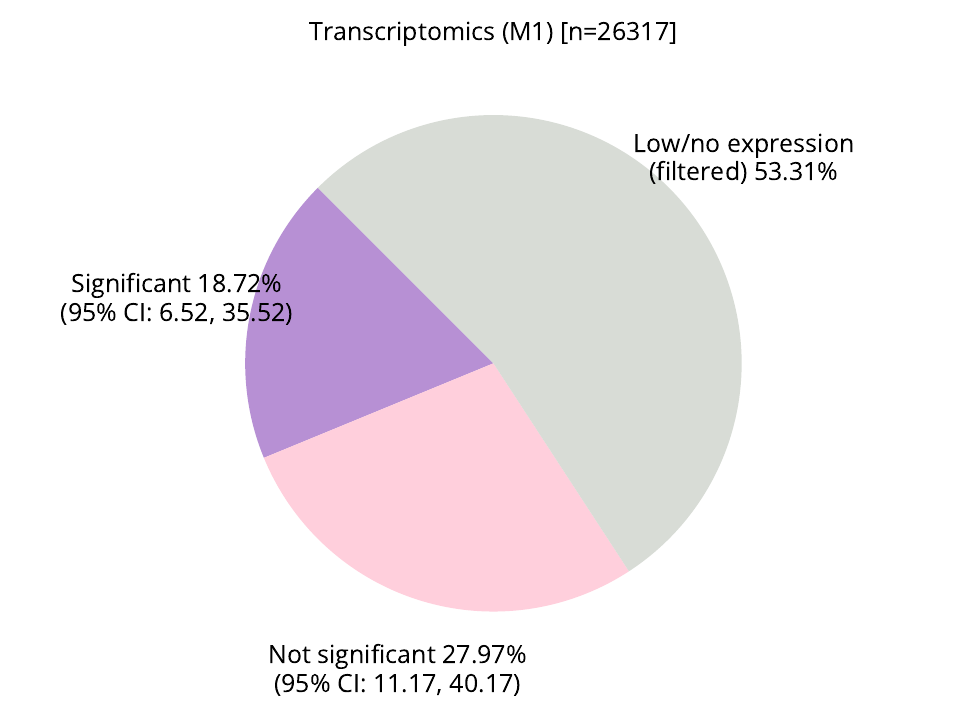}
    \includegraphics[width=0.48\textwidth]{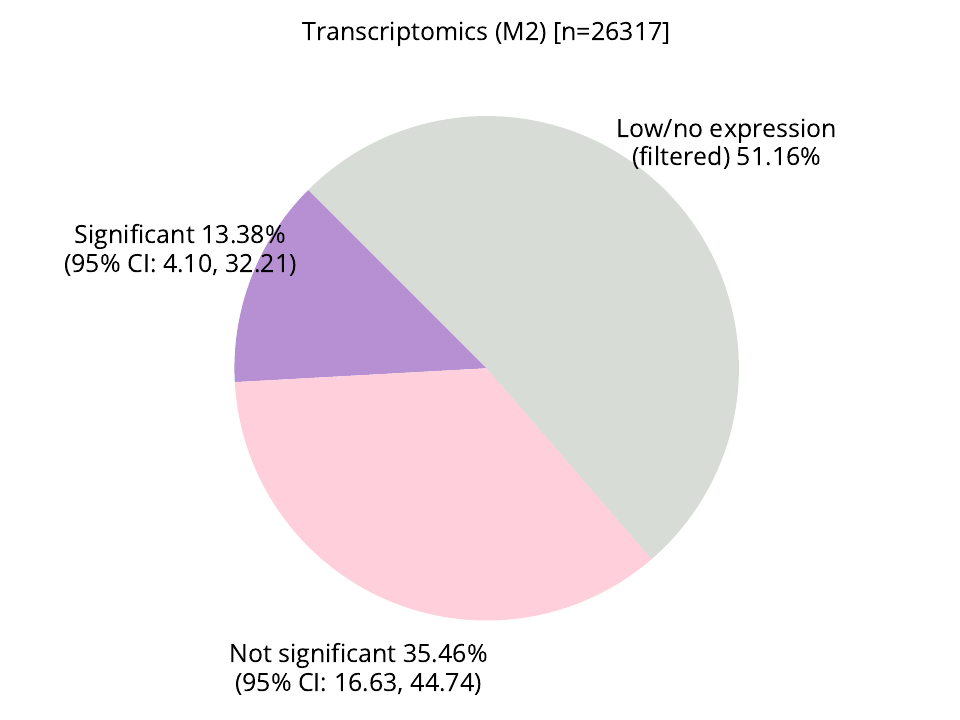}
    \includegraphics[width=0.48\textwidth]{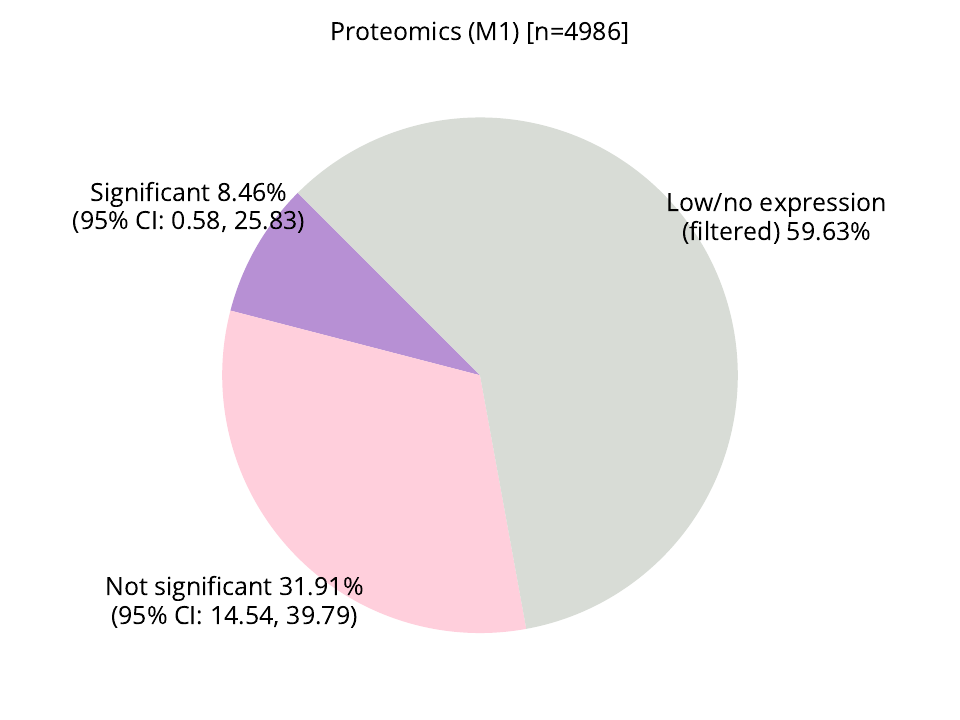}
    \includegraphics[width=0.48\textwidth]{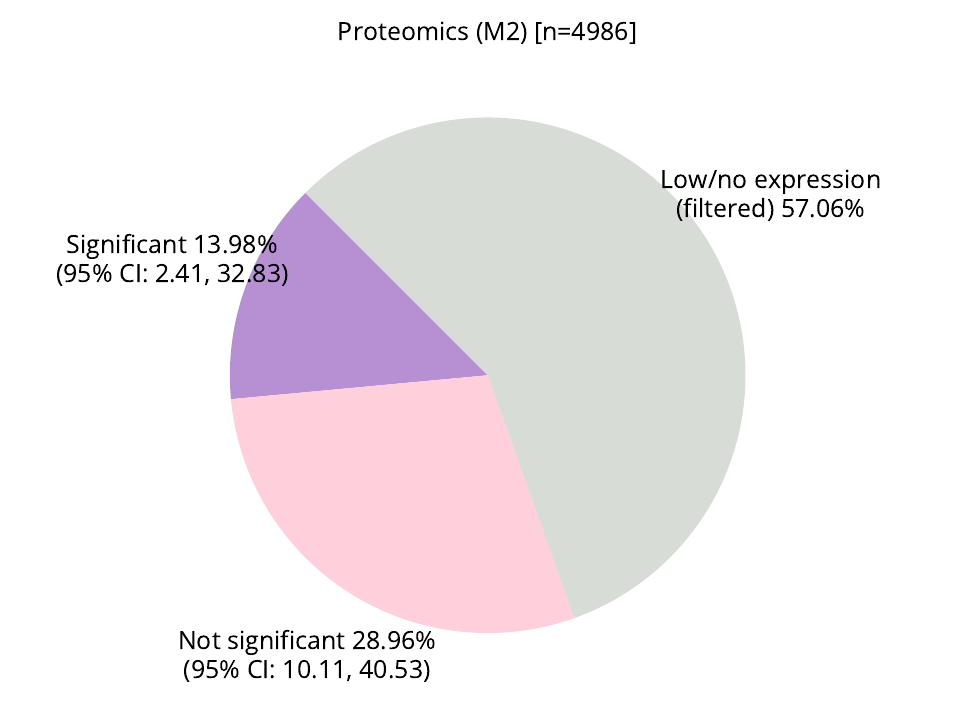}
    \caption{\textbf{Predictability across gene product types and stimuli.} Charts displaying the percentage of transcripts (top) and proteins (bottom) in M1 (left column) and M2 (right column) polarised states that are significantly ($p<0.05$) more predictable (SMAPE$_\text{\themethod}$ $<$ SMAPE$_\text{mean}$) from image data on held-out test data using \themethod{} than using the mean observed abundance in the training set (purple), the percentage of non significant marker proteins and transcripts (light pink) and those filtered out due to low or no observed expression in the experiment (light grey). \correctednn{We found that overall transcriptomics and proteomics were similarly frequently predictable with the exception of the M1 state in which transcriptomics abundances were more frequently predictable than proteomics abundances.}}
           \label{fig:pie_charts}
\end{figure}

\paragraph{Predictability by gene product.} We evaluated the degree to which the abundance of each gene product (transcripts and proteins) can be predicted directly from cellular images by comparing the prediction performance of \themethod{} to the Mean Predictor baseline. Mean Predictor is defined as a model that predicts the mean abundance value of the same gene product in the training data irrespective of inputs. We calculated the $r^2$ value between the ground truth and predicted abundances of gene products in the test set to determine the magnitude of the predictability improvement of the gene product abundances stemming from the use of the cellular imaging data (\Cref{fig:overall_performances}). In addition, for each gene product, we performed a one-sided Mann-Whitney-Wilcoxon test with $\alpha = 0.05$ to determine whether \themethod{} produces a statistically significant improvement in predictability over the Mean Predictor (\Cref{fig:pie_charts}). Out of all gene products that met minimal expression thresholds, we found that gene products are significantly predictable from cellular images in respectively \correctednn{18.72\% (95\% CI: 6.52\%, 35.52\% and 13.38\% (95\% CI: 4.10\%, 32.21\%)} of cases for transcripts in M1 and M2 polarised macrophages and \correctednn{8.46\% (95\% CI: 0.58\%, 25.83\%) and 13.98\% (95\% CI: 2.41\%, 32.83\%)} of cases for proteins in M1 and M2 polarised macrophages (\Cref{fig:pie_charts}). A selection of the top and bottom 10 gene products with the respectively lowest and highest $r^2$ values and symmetric mean average percentage errors (SMAPEs) for predicting gene product abundances achieved by \themethod{} is presented in \Cref{tb:top_genes_performances_1,tb:top_genes_performances_2,tb:top_genes_performances_3,tb:top_genes_performances_4}. The associations between the actual measured abundances and the predicted abundances based on the imaging data for gene products with the highest and lowest predictability are shown in \Cref{fig:scatter_plots_between_gene_products_and_predictability}.

\begin{figure}
    \centering
        \includegraphics[width=0.46\textwidth]{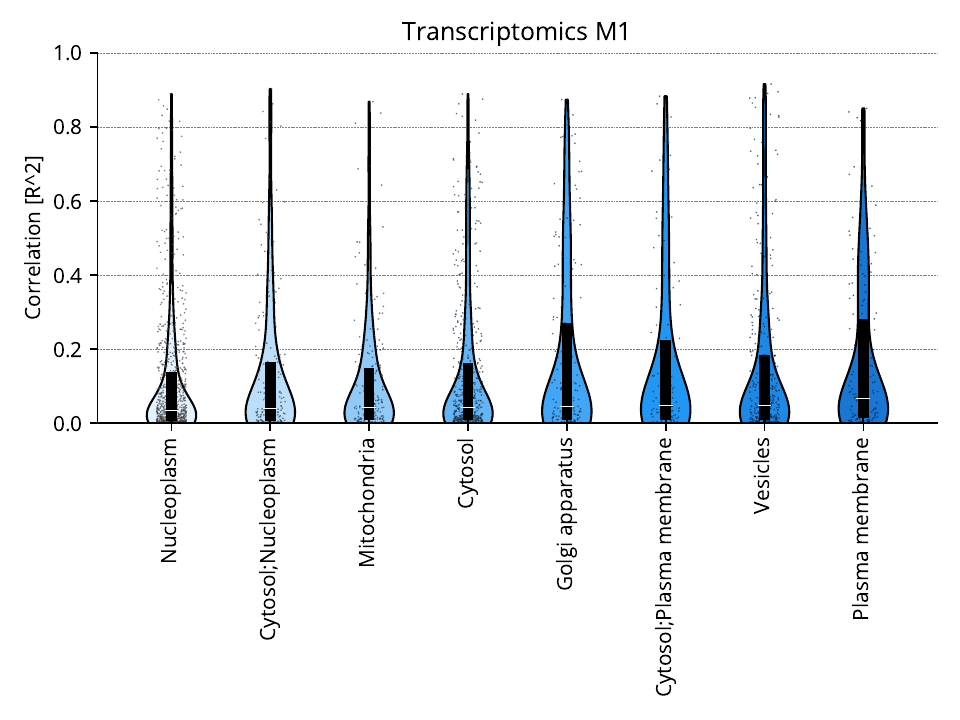}
        \includegraphics[width=0.46\textwidth]{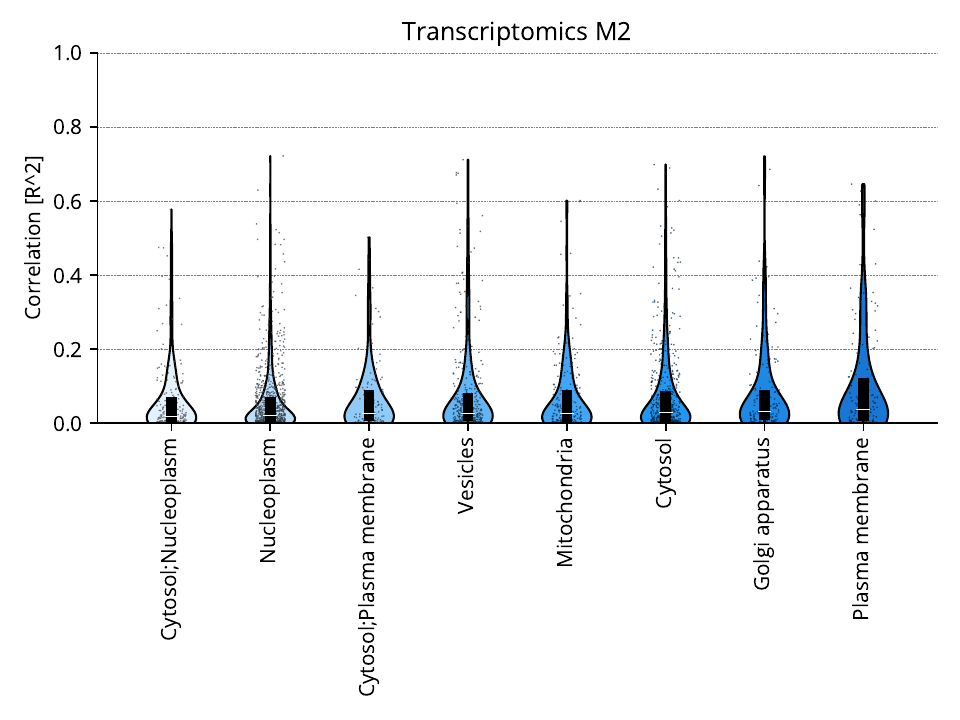}
        \includegraphics[width=0.46\textwidth]{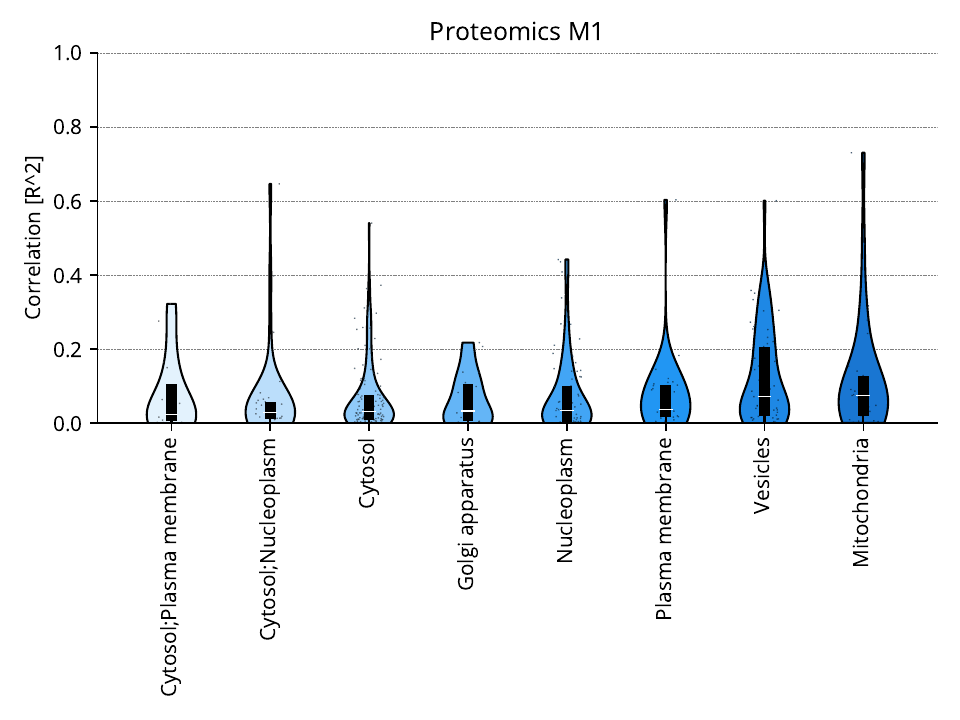}
        \includegraphics[width=0.46\textwidth]{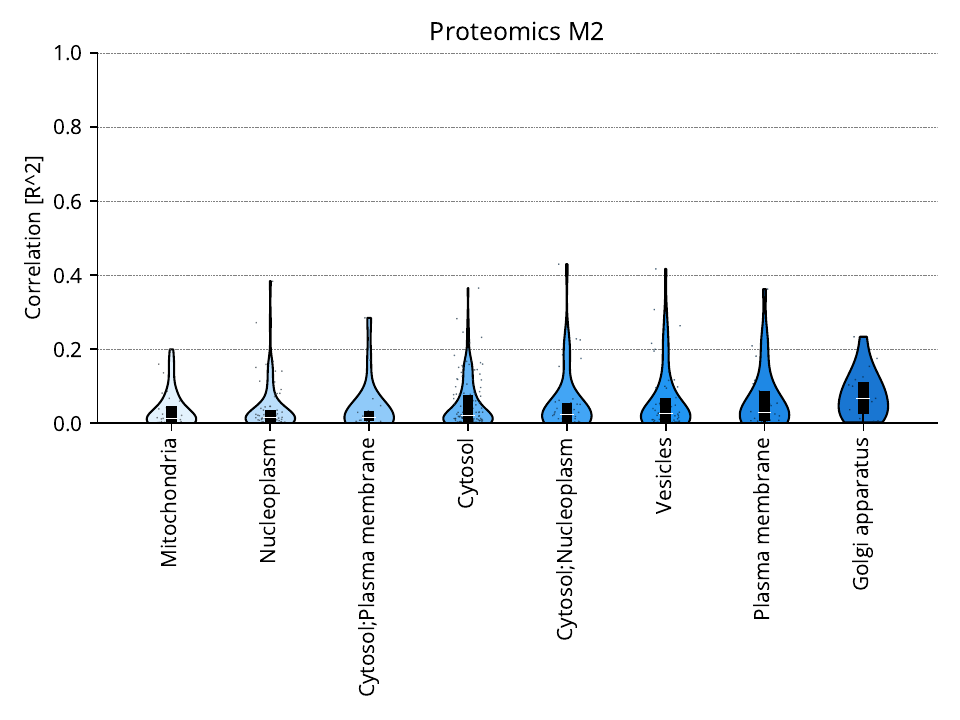}
       \caption{\textbf{Predictability by subcellular localisation of gene products.} Performance in predicting measured transcript (top) and protein (bottom) abundances from image data alone measured in terms of correlation (measured in $r^2$; y-axis, {higher} is better) on held-out test images of M1 (left column) and M2 (right column) macrophages broken down by subcellular location (x-axis) sorted from {least} (left) to {best} (right) predictable on average for the top 8 largest subcellular localisation categories. \correctednn{We found that gene products with subcellular locations in plasma membrane and vesicles are more predictable than other subcellular locations - with some exceptions (e.g. mitochondria being more predictable in the M1 state proteomics measurements)}. Note that gene products without known subcellular location are not shown here. Performance is calculated over all perturbed and unperturbed cell states to cover a diverse range of cellular states.
       }
       \label{fig:subcellular_localisation}
\end{figure}

\paragraph{Predictability by subcellular localisation.} We also evaluated the prediction error measured by $r^2$ broken down by subcellular localisation of the gene product as catalogued in the Human Protein Atlas (\url{https://www.proteinatlas.org/}) \cite{thul2017subcellular}. \correctednn{We observed that gene products localised in plasma membrane and vesicles were overall associated with higher $r^2$ compared to other subcellular localisations -- a pattern that held with some exceptions (e.g. M1 proteomics) across both protein and transcript abundances as well as macrophage polarisation states (\Cref{fig:subcellular_localisation})}.

\begin{figure}
    \centering
        \hspace*{-6.75em}\includegraphics[width=0.8\textwidth]{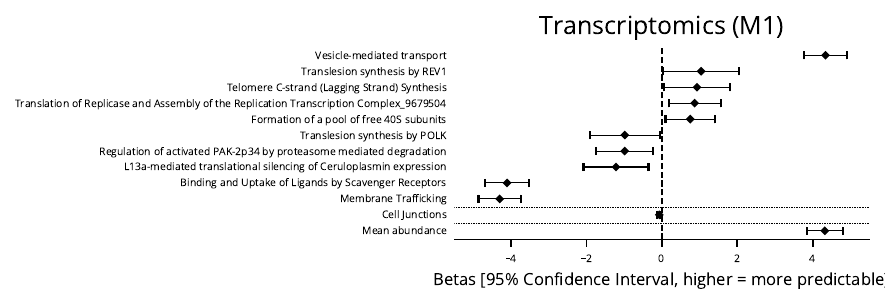}
        \hspace*{-4.0em}\includegraphics[width=0.8\textwidth]{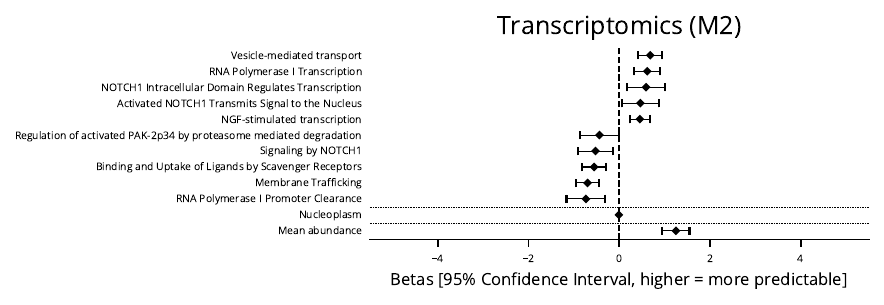}
        \hspace*{-1.8em}\includegraphics[width=0.8\textwidth]{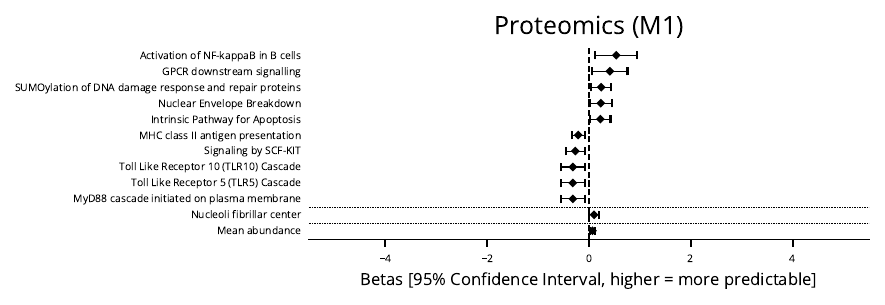}
        \hspace*{-3.2em}\includegraphics[width=0.8\textwidth]{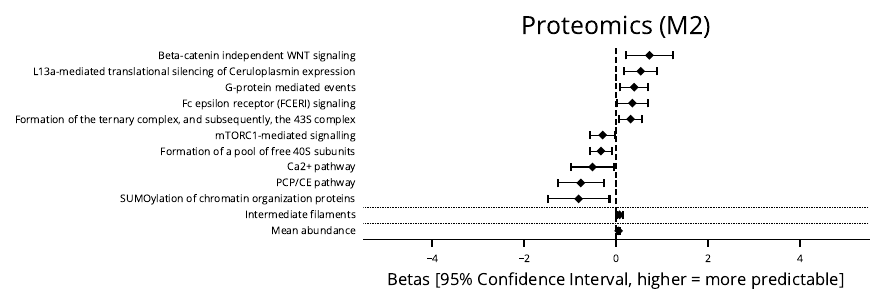}

       \caption{\textbf{Features associated with predictability of gene products.} Forest plots indicating the associations (y axis) with a significantly ($p<0.05$) predictability of protein and transcript abundances on held-out test data (x-axis; measured in linear regression beta coefficients; higher = more predictable) in M1 and M2 conditions including pathway membership of the predicted gene (first section from the top; top 5 pathways associated with lower and higher predictability pathways shown), sub-cellular localisation of the gene (second section from the top), and mean abundance levels observed in the training set (bottom-most section). We found that abundances are more predictable on average for more highly expressed protein products, and that membership in a variety of pathways is associated with differences in predictability of abundances.
}
       \label{fig:forest_plot}
\end{figure}

\paragraph{Predictability by pathway membership, number of modes of the measurement distribution, and abundance.} We also evaluated whether and to what degree there are linear associations between select properties of genes (pathway membership, localisation and abundance levels) and their predictability as measured by the $r^2$ value between \themethod{} predicted and experimentally observed test set abundance levels. We observed that abundances are more predictable on average for more highly expressed protein products, and that membership in a variety of pathways is associated with differences in predictability of protein and transcript abundances 
(\Cref{fig:forest_plot}). 

\paragraph{Qualitative analysis of feature importance.} To better understand the factors associated with better prediction performance for individual gene products, we qualitatively investigated the cellular images corresponding to a selection of the abundance predictions, the prediction performance of \themethod{} for those images, and feature attribution heatmaps that indicate areas of focus for the model 
\corrected{(Figure 7)}. \corrected{Feature attribution heatmaps are calculated using SmoothGrad, a gradient-based explanation method that enhances heatmaps by removing irrelevant noisy regions \cite{smilkov2017smoothgrad}.}
By visual analysis, we found that \themethod{} focused on the general cell shape with an apparent enhanced emphasis on the nucleus - a finding that was robust across perturbations, stimulation conditions and predicted omics marker type. In addition, we found that the generated attribution maps were distributing weight across larger parts of the image in image tiles that contained multiple cells in addition to the cell that the tile was centred on, which indicates that \themethod{} in some cases attempted to leverage information distributed across cells in a single tile.

\paragraph{Cell image embedding.} In addition, we evaluated the ability of the image-based feature embedding learnt by the image-embedding component of \themethod{} to differentiate between perturbations and functional states of cell populations (\Cref{fig:embedding}). We found that the image-based feature embedding learnt by \themethod{} is able to differentiate between various cell conditions and captures rich information on cell function and state -- implying that its ability to predict omics markers is closely linked to its capability to accurately identify cell states and conditions.

\begin{figure}
\vspace{-4em}
    \centering

       \begin{subfigure}[b]{0.37\textwidth}
        \centering
        \raisebox{0.0em}{\hspace{5.2em}Predictions}
    \end{subfigure}
    \begin{subfigure}[b]{0.3\textwidth}
        \raisebox{0.0em}{\hspace{3.1em}Patch}
    \end{subfigure}
    \begin{subfigure}[b]{0.3\textwidth}
        \raisebox{0.0em}{\hspace{-3.15em}Attribution}
    \end{subfigure}

    \begin{subfigure}[b]{0.03\textwidth}
        \centering
        \raisebox{4.5em}{\rotatebox[origin=t]{90}{PDCD6IP Protein (M1)}}
    \end{subfigure}
    \begin{subfigure}[b]{0.03\textwidth}
        \centering
        \raisebox{4.5em}{\rotatebox[origin=t]{90}{weakly predictable}}
    \end{subfigure}
    \centering
    \begin{subfigure}[b]{0.31\textwidth}
        \centering
    \vspace{1em}
        \includegraphics[width=\textwidth]{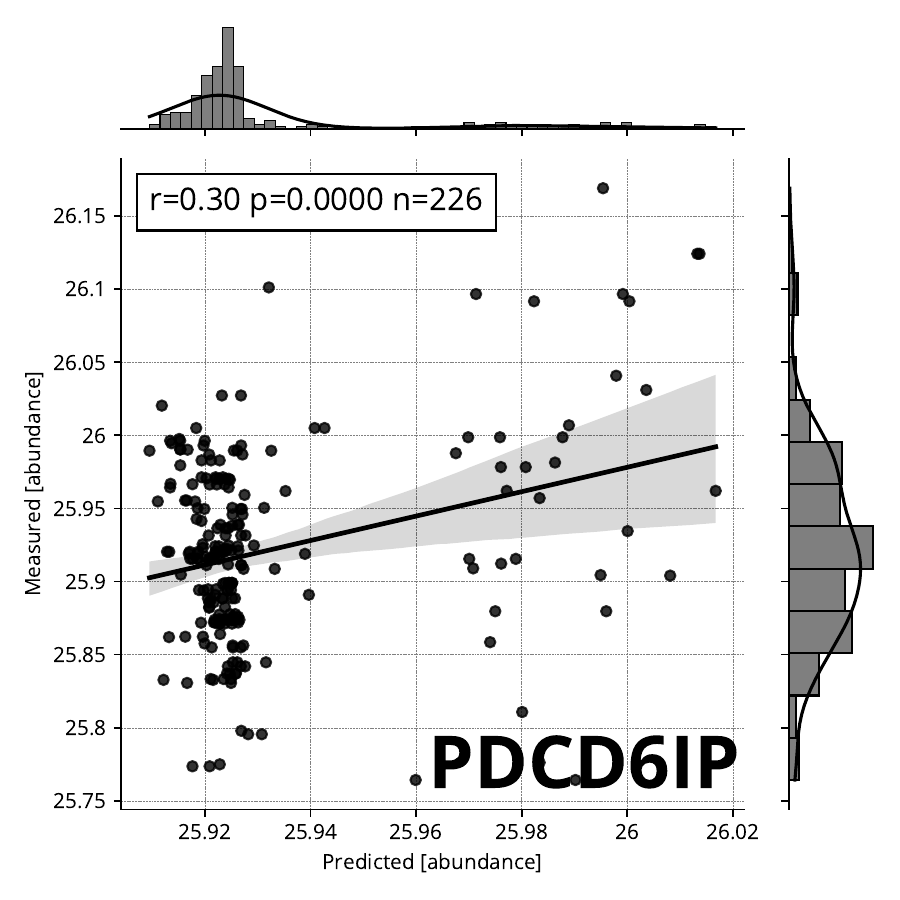}
    \end{subfigure}
    \centering
    \begin{subfigure}[b]{0.53\textwidth}
    \begin{subfigure}[b]{0.6\textwidth}
    \begin{subfigure}[b]{0.48\textwidth}
        \centering
        \includegraphics[width=\textwidth]{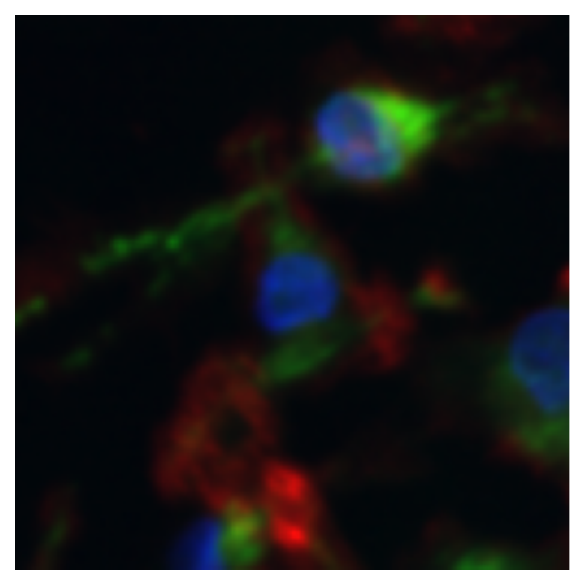}
    \end{subfigure}
    \begin{subfigure}[b]{0.48\textwidth}
        \centering
        \includegraphics[width=\textwidth]{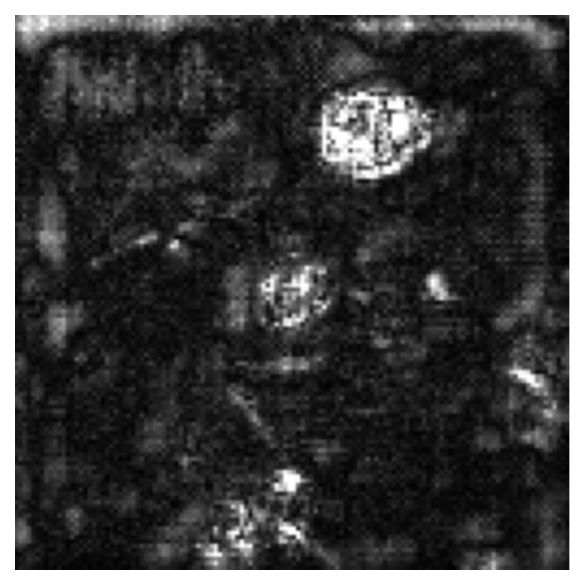}
    \end{subfigure}
    \begin{subfigure}[b]{0.48\textwidth}
        \centering
        \includegraphics[width=\textwidth]{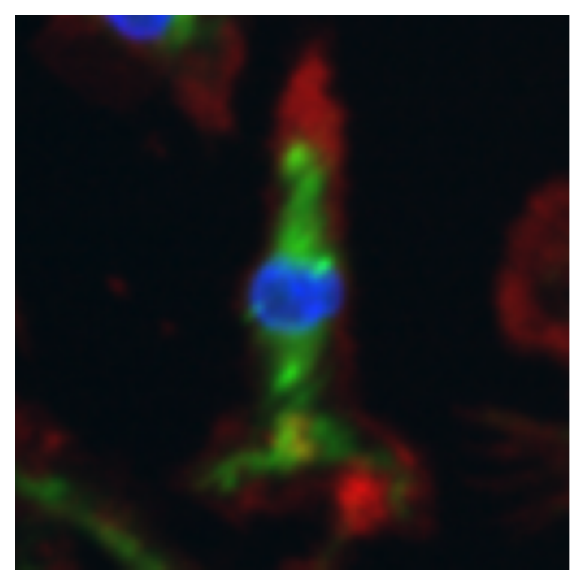}
    \end{subfigure}
    \begin{subfigure}[b]{0.48\textwidth}
        \centering
        \includegraphics[width=\textwidth]{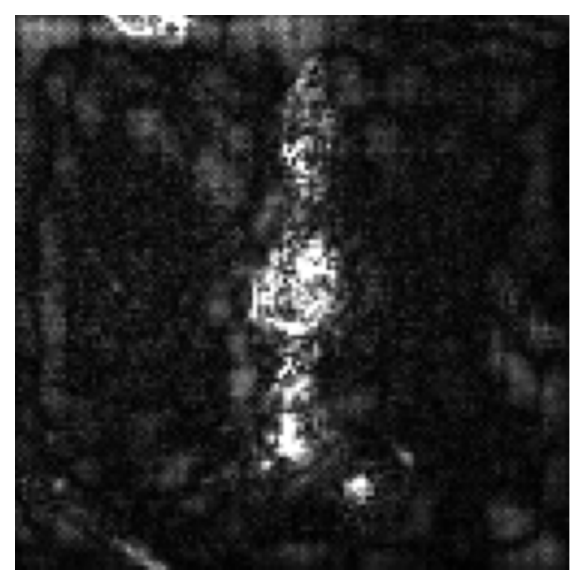}
    \end{subfigure}
    \end{subfigure}
    \end{subfigure}

    \begin{subfigure}[b]{0.03\textwidth}
        \centering
        \raisebox{4.5em}{\rotatebox[origin=t]{90}{TFPI2 Transcript (M1)}}
    \end{subfigure}
    \begin{subfigure}[b]{0.03\textwidth}
        \centering
        \raisebox{4.5em}{\rotatebox[origin=t]{90}{poorly predictable}}
    \end{subfigure}
    \centering
    \begin{subfigure}[b]{0.31\textwidth}
        \centering
    \vspace{1em}
        \includegraphics[width=\textwidth]{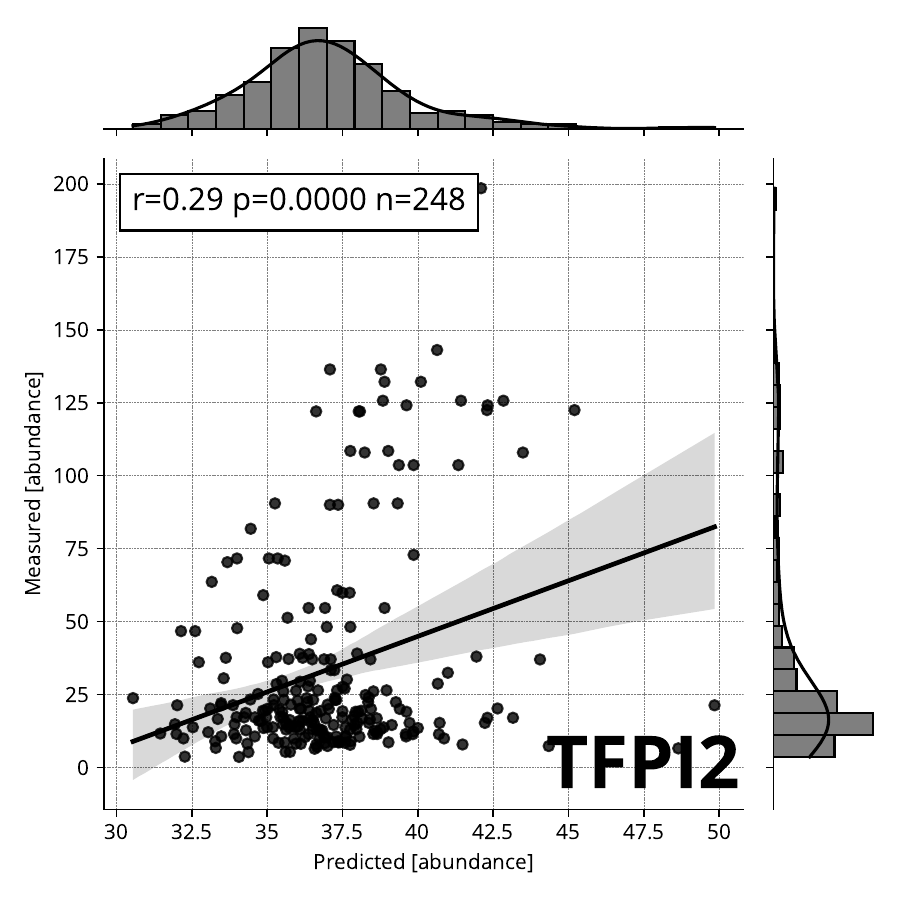}
    \end{subfigure}
    \centering
    \begin{subfigure}[b]{0.53\textwidth}
    \begin{subfigure}[b]{0.6\textwidth}
    \begin{subfigure}[b]{0.48\textwidth}
        \centering
        \includegraphics[width=\textwidth]{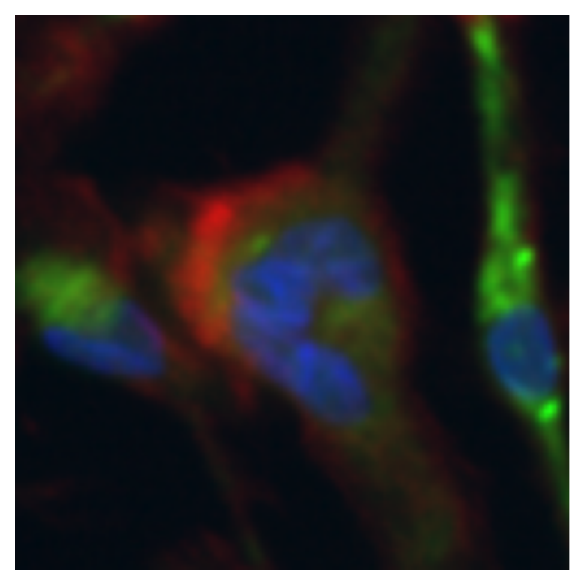}
    \end{subfigure}
    \begin{subfigure}[b]{0.48\textwidth}
        \centering
        \includegraphics[width=\textwidth]{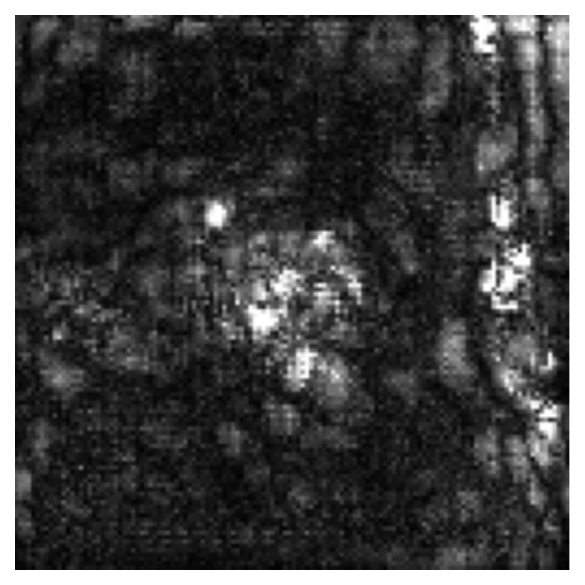}
    \end{subfigure}
    \begin{subfigure}[b]{0.48\textwidth}
        \centering
        \includegraphics[width=\textwidth]{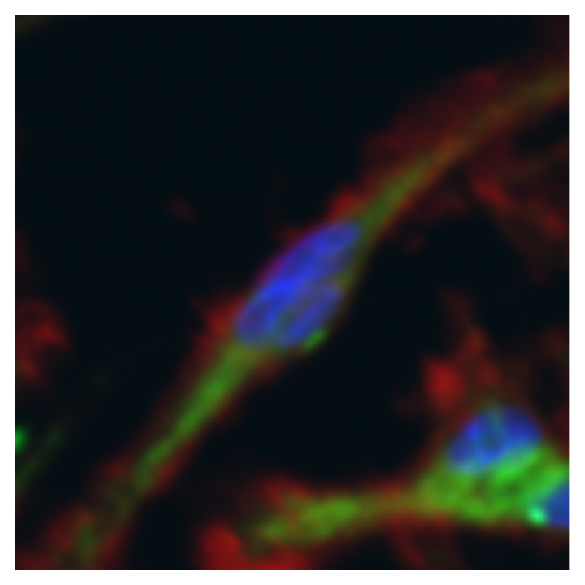}
    \end{subfigure}
    \begin{subfigure}[b]{0.48\textwidth}
        \centering
        \includegraphics[width=\textwidth]{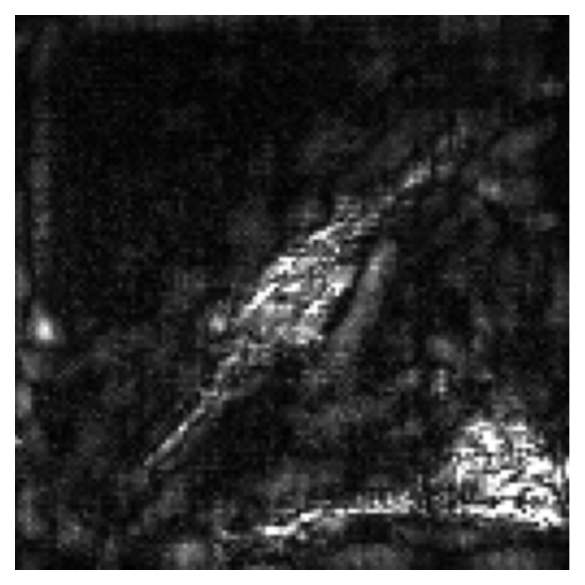}
    \end{subfigure}
    \end{subfigure}
    \end{subfigure}
    \vspace{1em}

    \begin{subfigure}[b]{0.03\textwidth}
        \centering
        \raisebox{4.5em}{\rotatebox[origin=t]{90}{DDI2 Protein (M1)}}
    \end{subfigure}
    \begin{subfigure}[b]{0.03\textwidth}
        \centering
        \raisebox{4.5em}{\rotatebox[origin=t]{90}{not predictable}}
    \end{subfigure}
    \centering
    \begin{subfigure}[b]{0.31\textwidth}
        \centering
        \includegraphics[width=\textwidth]{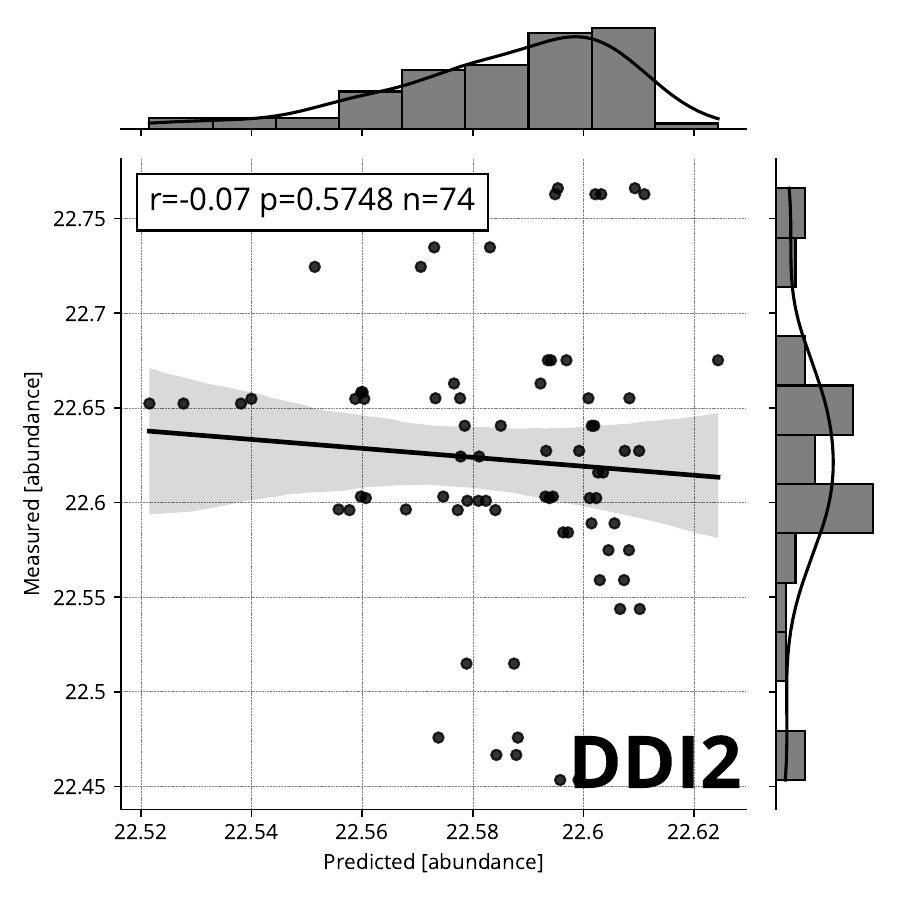}
    \end{subfigure}
    \centering
    \begin{subfigure}[b]{0.53\textwidth}
    \begin{subfigure}[b]{0.6\textwidth}
    \begin{subfigure}[b]{0.48\textwidth}
        \centering
        \includegraphics[width=\textwidth]{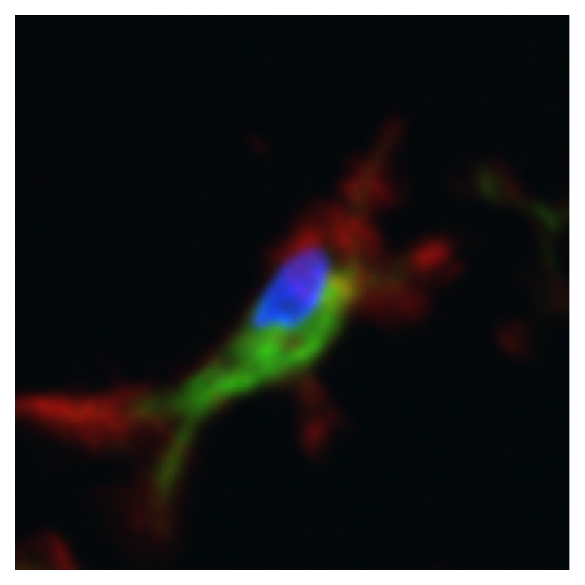}
    \end{subfigure}
    \begin{subfigure}[b]{0.48\textwidth}
        \centering
        \includegraphics[width=\textwidth]{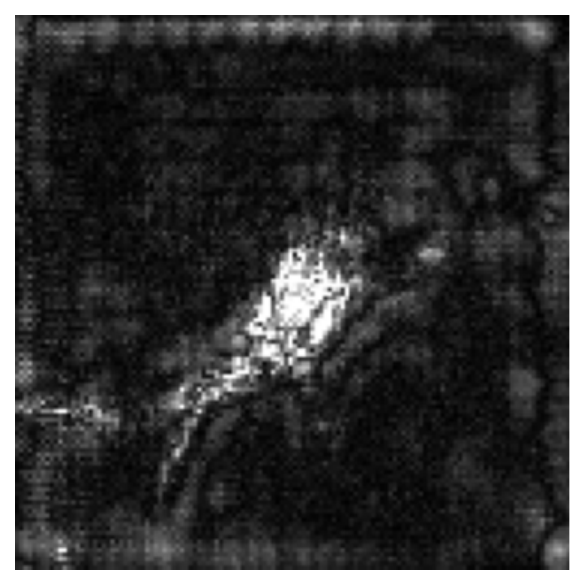}
    \end{subfigure}
    \begin{subfigure}[b]{0.48\textwidth}
        \centering
        \includegraphics[width=\textwidth]{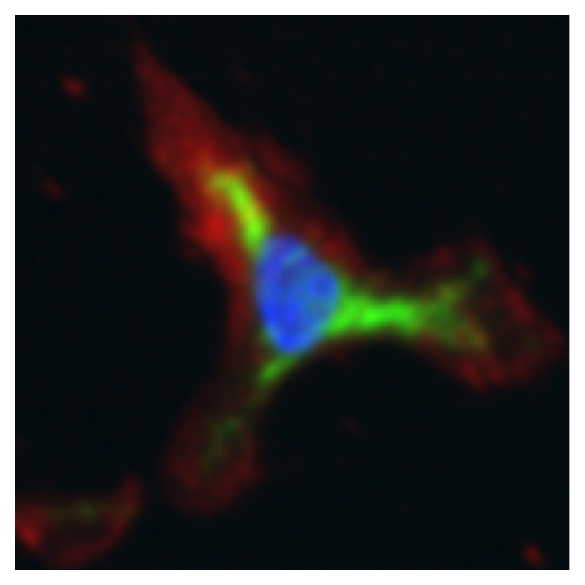}
    \end{subfigure}
    \begin{subfigure}[b]{0.48\textwidth}
        \centering
        \includegraphics[width=\textwidth]{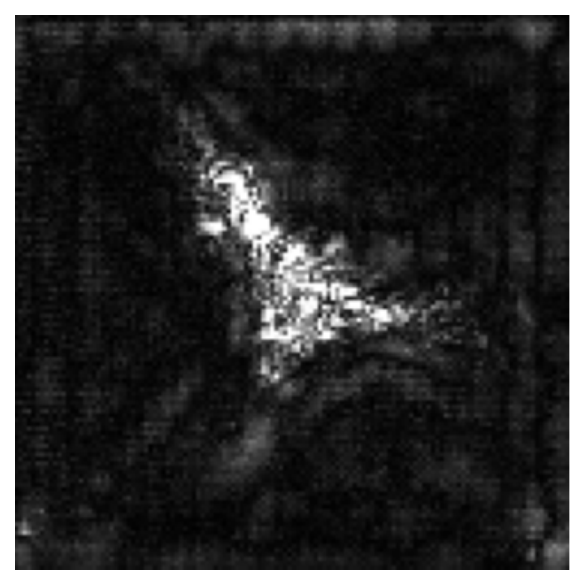}
    \end{subfigure}
    \end{subfigure}
    \end{subfigure}

    \begin{subfigure}[b]{0.03\textwidth}
        \centering
        \raisebox{4.5em}{\rotatebox[origin=t]{90}{HLA-A Transcript (M2)}}
    \end{subfigure}
    \begin{subfigure}[b]{0.03\textwidth}
        \centering
        \raisebox{4.5em}{\rotatebox[origin=t]{90}{decently predictable}}
    \end{subfigure}
    \centering
    \begin{subfigure}[b]{0.31\textwidth}
        \centering
    \vspace{1em}
        \includegraphics[width=\textwidth]{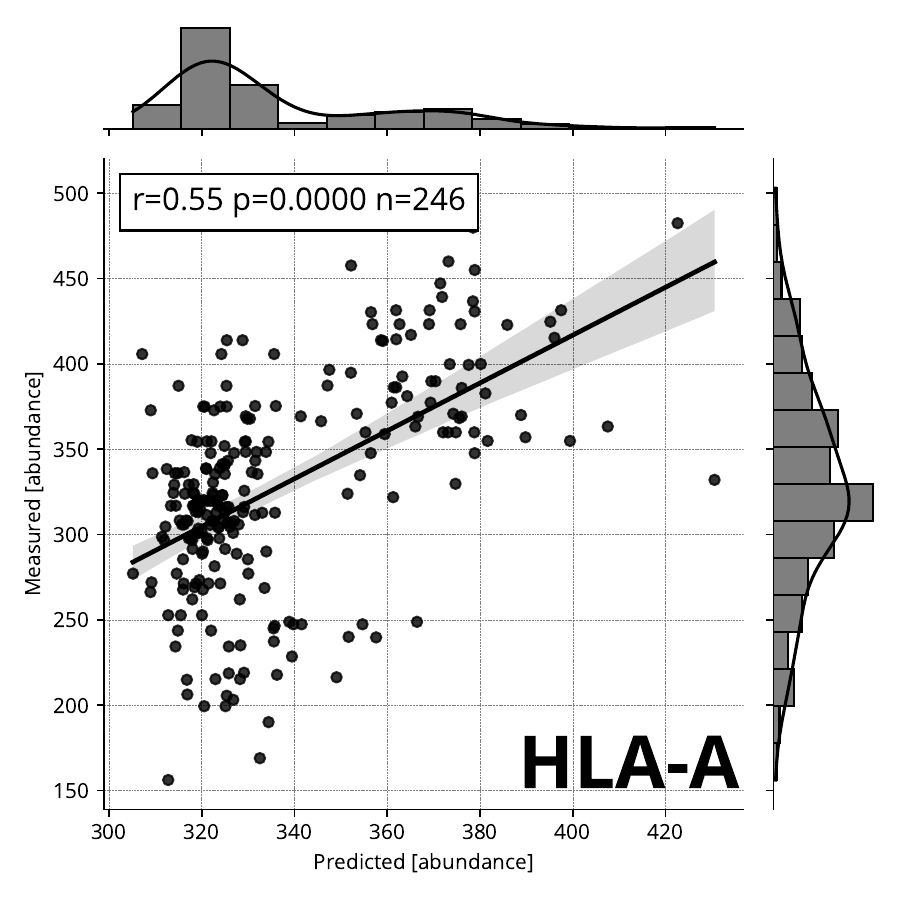}
    \end{subfigure}
    \centering
    \begin{subfigure}[b]{0.53\textwidth}
    \begin{subfigure}[b]{0.6\textwidth}
    \begin{subfigure}[b]{0.48\textwidth}
        \centering
        \includegraphics[width=\textwidth]{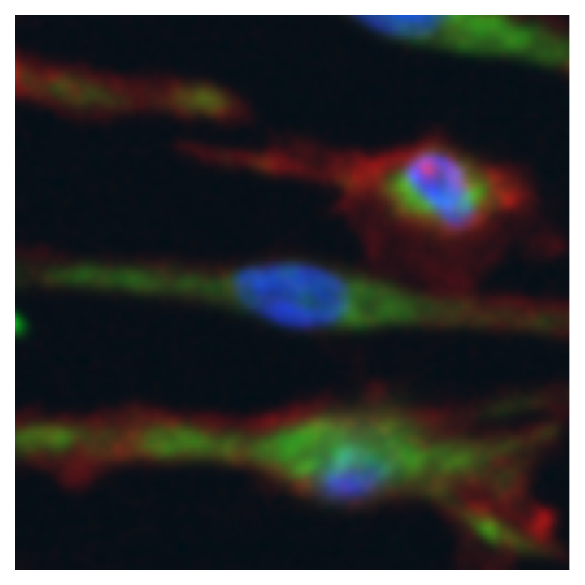}
    \end{subfigure}
    \begin{subfigure}[b]{0.48\textwidth}
        \centering
        \includegraphics[width=\textwidth]{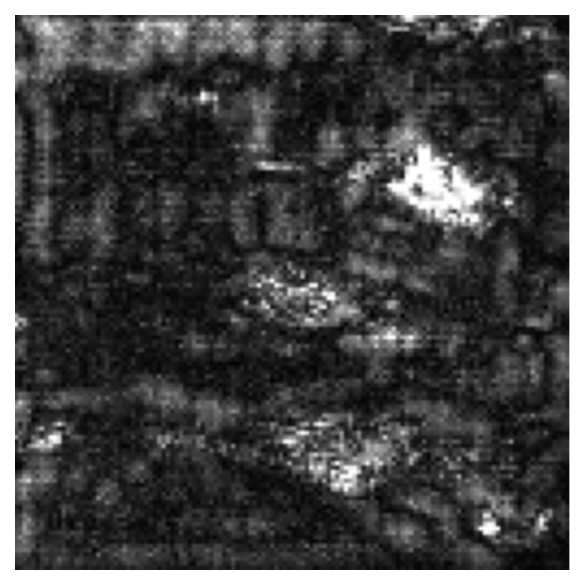}
    \end{subfigure}
    \begin{subfigure}[b]{0.48\textwidth}
        \centering
        \includegraphics[width=\textwidth]{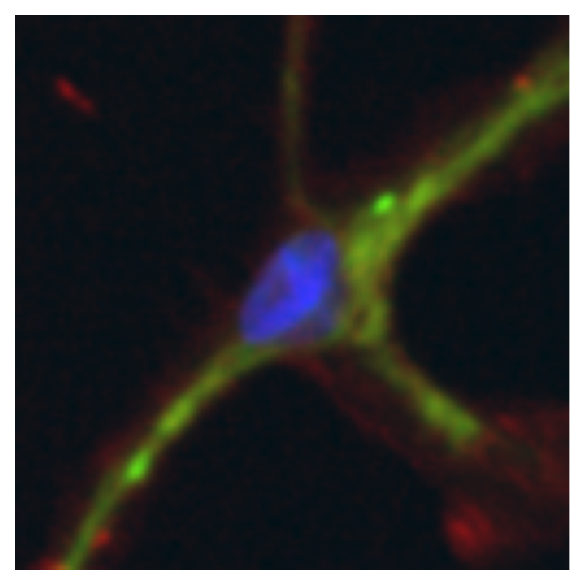}
    \end{subfigure}
    \begin{subfigure}[b]{0.48\textwidth}
        \centering
        \includegraphics[width=\textwidth]{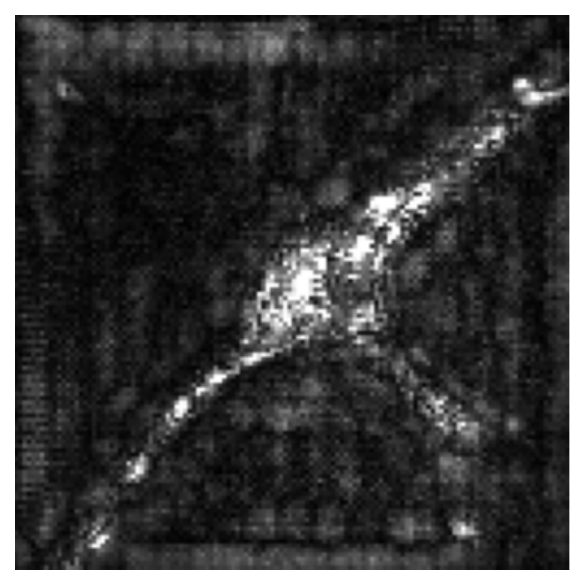}
    \end{subfigure}
    \end{subfigure}
    \end{subfigure}
       \caption{\textbf{Example gene products.} Selected gene products (from top to bottom: PDCD6IP, TFPI2, DDI2, HLA-A) and the associations between measured and predicted abundances based on the imaging data from the held-out test fold (leftmost column; \corrected{each point represents a well in the dataset}), sample cell-centred image patches (centre column; blue = Hoechst, green = MitoTracker, red = phalloidin) and the assigned attributions that visualise the importance assigned to pixels in the original image by the model to make its prediction (rightmost column). It is important to consider the performance of the predictive model when interpreting attribution maps.  \corrected{The number of wells per product may differ due to exclusion of wells with low/no gene product counts (\Cref{sec:method}).}}
       \label{fig:selected_genes}
\end{figure}

\begin{figure}
        \centering
        \includegraphics[width=\textwidth]{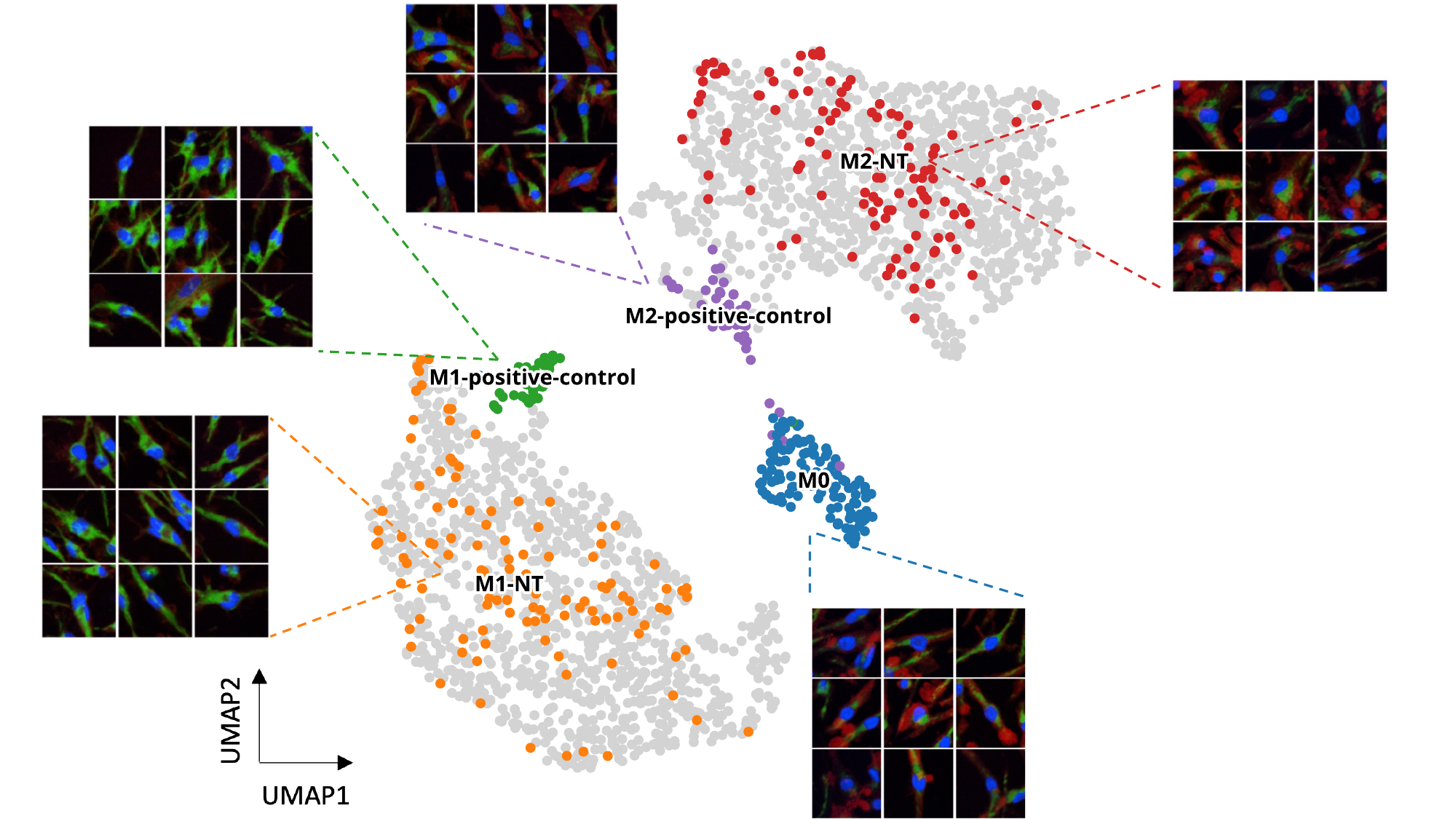}
       \caption{\textbf{Image embeddings recapitulate cell function and state.} Learned morphological feature embeddings in a 2-dimensional Uniform Manifold Approximation, Projection for Dimension Reduction (UMAP) \cite{becht2019dimensionality} space as derived from high-content cell imaging. \corrected{Each point represents a well and the colored points are wells that have known reference states for macrophage biology (NT = non targeted; M1 positive control = STAT1 = M1 reverting; M2 positive control = STAT6 = M2 reverting). For each of these reference states, there are associated image insets with 9 cell-centered patches that were randomly selected to represent that well.} The learned morphological embedding is well able to distinguish between selected genetic perturbations that have emphasised morphological effects (other colors with callouts) - indicating that morphological features recapitulate differences in cell state as intended by the model training (\Cref{sec:imageanalysis}) and that the cell imaging embeddings carry rich information on cell function and state. Plate batch effects were removed using the CORrelation ALignment (CORAL) algorithm \cite{sun2016return}.}
        \label{fig:embedding}
\end{figure}

\section{Discussion}
\label{sec:discussion}

The ability to predict omics markers directly from cellular images could be of high utility for applications in cell-based assays where the use of omics technologies to measure molecular markers may not always be technically feasible or cost-effective in terms of time and resources. In contrast, high-content cellular imaging is a scalable and cost-effective alternative that is amenable to high-throughput and automated experimentation. As demonstrated in this study, high-content imaging data contains rich information on cellular state and function that is at least partially overlapping with that generated via omics readouts. In addition, unlike measurement protocols for omics technologies that require the dissolution of cells (such as, e.g., RNAseq), cellular imaging has the additional advantage of being a non-destructive method for obtaining protein and transcript abundance measurements from cells and could thus {with the substitution of live- rather than fixed-cell labels as used herein} enable the tracking of omics marker evolution in the same cell population over time. Image-based prediction of omics markers could have particular utility for iterative experimental exploration campaigns \cite{mehrjou2021genedisco,lyle2023discobax} in a fixed set of conditions and cell types where sufficient training data with paired omics and imaging are available. Given the shared information content in cellular images and omics measurements, potential avenues for future work could be in exploring the inverse task of generating images from multi-omics measurements \cite{Lee2022morphnet,dumont2021overcoming}. It is important to note that the predictions of \themethod{} are associational based on observed patterns in its training data, and -- although interpreting models such as \themethod{} can help us better understand the interplay between cell morphology and molecular measurements -- uncovering the causal molecular mechanisms \cite{qiu2020inferring,scherrer2021activecausal,chevalley2022causalbench,chevalley2023causalbench} that give rise to changes in cellular morphology and gene product localisation (and vice-versa) remain open challenges.

The ability of \themethod{} to predict omics measurements across a wide range of gene products from cellular imaging data implies that at least some of the information on cellular state and function contained in cellular images is overlapping with that described by omics measurements. The results presented in this study therefore demonstrate that gene products' proteomics and transcriptomics measurements can indeed be predicted with significantly better performance with cellular imaging data than in its absence, with considerable prediction performance in many cases -- as implied by the known interplay between cellular morphology and protein localisation and molecular biology \cite{wada2011hippo,hamilton2001regulation,canton2006shape,tsuji2018morphology}. However, for these predictions to reach practical utility, it is likely necessary to reach a certain minimal performance threshold for predicting a defined set of markers that depends on the use-case of interest. Given that not all gene products are strongly predictable from cellular images and that certain categories of markers appear to be on average associated with higher prediction errors (for example, cytosol-localised gene products as shown in \Cref{fig:subcellular_localisation}), imaging-based omics prediction by \themethod{} cannot be considered a general substitute of omics measurements. The measured prediction performance for the gene products of interest in the cellular population of interest can, however, serve as a guideline to which omics layers for which gene products could potentially be substituted by predictions based on cellular imaging data. {Beyond prediction performance, we note that \themethod{} - although trained on bulk measurements - produces omics predictions from sets of image tiles and therefore could in the future be used to interrogate heterogeneity among cell populations or even single cells.} Furthermore, we found that predicted and observed gene product abundances are sometimes bi- or multimodal in their distributions - potentially reflecting the underlying cell states associated with different levels of gene product abundances (\Cref{fig:selected_genes}).

\paragraph{Limitations.} A major limitation of the \themethod{} approach to predicting omics measurements directly from cellular images is that sufficient training data consisting of paired omics and cellular imaging data for a cell population of interest, or alternatively a pre-trained model initialised with such training data, are required. Similar to other machine-learning domains with shared commons (for example, the HuggingFace model repository (\url{https://huggingface.co/models})), a potential approach to combat this limitation could be to create a central resource of shared pre-trained omics prediction models for a variety of cell types that could enable researchers to leverage the previous experimental work of others. A potential barrier to a collaborative effort to generate shared pre-trained \themethod{} models could be the difficulty of standardising and pooling imaging data collected under different experimental configurations and using different equipment. In addition, although the presented experimental evaluation in this work considered a wide range of extreme cell states induced by genetic perturbations and different experimental stimulations, additional experiments are needed to substantiate the robustness of \themethod{} predictions - including testing across different donors, in single cells rather than batches and across a wider range of stimuli, perturbations and cell types than presented in this study, so that the predictability of various omics modalities across biological and experimental contexts can be fully determined. 

In summary, we developed \themethod{} -- a new \corrected{application of} deep learning methods for predicting bulk transcriptomics and proteomics directly from cell images. We demonstrated that image information is predictive of multi-omics for  \correctednn{4927 (18.72\%; 95\% CI: 6.52\%, 35.52\%) and 3521 (13.38\%; 95\% CI: 4.10\%, 32.21\%)} transcripts out of 26137 in M1 and M2-stimulated macrophages respectively and for \correctednn{422 (8.46\%; 95\% CI: 0.58\%, 25.83\%) and 697 (13.98\%; 95\% CI: 2.41\%, 32.83\%)} proteins out of 4986 in M1 and M2-stimulated macrophages respectively. To the best of our knowledge, this is the first study that quantifies the predictability of transcriptomics and proteomics from high-content cellular images across a wide range of cellular states and under perturbations. Our results imply that high-throughput cellular imaging assays could, in some settings and depending on the mechanisms of interest, be a scalable and cost-effective alternative to direct multi-omics measurements, and also highlight a rich interplay between information contained in cellular images and the underlying molecular state of a cell population. We believe our results warrant future studies on the predictability of additional multi-omics layers (for example, metabolomics and secretomics) from cellular images and on the predictability of multi-omics across additional cell types that are amenable to paired imaging and multi-omics workflows.

\section{Materials and methods} %
\label{sec:method}

\subsection{Data Acquisition}
\label{sec:data_acquisition}
A library of 156 genes were knocked out in induced pluripotent stem cell (iPSC) derived macrophages \corrected{from a single donor} using a high throughput CRISPR (clustered regularly interspaced short palindromic repeats)-Cas9 (CRISPR associated protein 9) method [41]. Macrophages were then cultured for recovery for 6 days in 384 well plates (PerkinElmer PhenoPlate™ 384-well microplates, cat. no. 6057302). The resulting cells were stimulated for 24 hours with \corrected{cytokines to induce proinflammatory (M1)-like or anti-inflammatory (M2)-like states.}

Plates were then processed for staining. Briefly, culture media (50 µL per well) was removed using a Bravo automated liquid handler (Agilent). 50 µL of live cell MitoTracker staining solution (MitoTracker orange (Invitrogen\uplett{TM}, cat. no. M7510) diluted in prewarmed (37 ºC) media for a final concentration of 200 nM) was immediately added into each well using a Multidrop (Multidrop Combi Reagent Dispenser, Thermo Scientific). Plates were incubated in dark for 30 minutes at 37 ºC, 95\% relative humidity and 5\% CO2, before MitoTracker staining solution was removed with Bravo. Cells were then fixed with 30 µL, 4\% paraformaldehyde (Alfa Aesar, cat. no. J61899) per well and left for 20 minutes at room temperature in dark. Plates were washed three times with PBS-tween (0.1\% v/v tween 20 (Sigma-Aldrich, cat. no. P9416) in Dulbecco’s Phosphate Buffered Saline (PBS, Sigma-Aldrich, cat. no. D8537)) using a plate washer (Biotek 405 TS Microplate washer, Agilent). Cells were permeabilized by adding 50µL permeabilization solution (0.1\% v/v Triton X-100 (Sigma-Aldrich, cat. no. T8787) in PBS) in to each well using a Multidrop at medium speed and incubated at room temperature (RT) for 20 minutes. After permeabilization solution was removed, plates were washed three times with PBS-tween using plate washer and 50µL of blocking solution (0.1\% v/v tween 20 and 2\% v/v goat serum (Biowest cat. no. S2000-500) in PBS), was added into each well using a Multidrop at a medium speed. Plates were sealed and left in dark at RT for 2 hours. Blocking solution was removed using plate washer and 30uL per well primary antibody mix in blocking solution was added \corrected{(antibody identity not relevant to the current study as it was included for alternate purposes)}. Plates were sealed and stored at 4 ºC in a dark fridge overnight. The following day, after washing with PBS-tween three times, 30 µL per well of a secondary labeling mixture containing 1:1000 nuclear stain (Hoechst 33342 Thermo H3570) and 1:400 phalloidin (Thermo cat. no. A22287) in blocking solution was added using a Multidrop at medium speed. Following 2 hours of incubation at RT, plates were washed with PBS-tween using the plate washer three times and left in 50uL PBS per well for imaging analysis.

\subsubsection{Library selection}
guide RNA (gRNA) targeting specific genes were selected for inclusion in the experiment based on prior evidence of the target gene affecting macrophage polarisation. The evidence used for gRNA prioritisation included genetic evidence in relevant immunological diseases (GWAS), expression data in monocytes and pathway diversity of included gRNAs. In selecting the gRNA library, we specifically aimed to cover a diverse representation of pathways to maximise likelihood of covering as much perturbation diversity as possible under the given experimental budget. 

\subsubsection{High-throughput 3’ RNA-seq}
The protocol was developed based on the DRUG-seq protocol \cite{li2022drugseq}. In brief, cells in 384 well plates were washed with ice-cold PBS on BRAVO (Agilent Technologies). 18ul lysis buffer (50mM Tris-HCL pH8.0, 75mM KCL, 6\%Ficoll PM-400, 0.15\%Triton-100, 0.5unit/ul Thermo Fisher SUPERaseIn) was added by Dragonfly (SPT Labtech) to lyse the cells directly in each well. Plates were sealed and shook at 900rpm for 10mins. To start the reverse transcription (RT) 1µl bar- coded RT primers at 10nM were added to 5µl RT reaction mix (1uM TSO 5’- AAGCAGTGGTATCAACGCAGAGTGAATrGrGrG-3’, 0.1mM dNTP, 0.2Mm GTP, 0.4unit/µl Thermo Fisher RNaseOUT, 2unit/µl Thermo Fisher Maxima H Minus Reverse Transcriptase, 1:20000 ERCC mix1) together with 15µl cell lysate. Plates were incubated at 42C for 90 mins followed by 10 cycles at 50C for 2 minutes, 42C for 2 minutes, together with a final incubation at 85C for 5 minutes. RT reaction mix from each well were pooled into a single sample purified with the Agencourt RNAClean XP beads (Beckman Coulter). cDNA was amplified using primer 5’- AAGCAGTGGTATCAACGCAGAGT-3’, beads purified, and tagmented using TDE1 enzyme and buffer kits (Illumina). Individ- ual libraries were indexed and sequenced on illumina Nova-seq 6000. Library was loaded at 2nM with read 1 at 20 cycles, index 1 at 8 cycles, index 2 at 0, and read 2 at 76 cycles.

\corrected{The mRNA reads (R2) were trimmed for poly-A and poly-G tails ($\geq$12nt) using Atropos and then aligned against the Human reference genome GRCh38 (Ensembl 96) using STAR. Read quantification (assignment to genes, demultiplexing of well barcodes and UMI-aware counting) was performed along with the alignment using the extension STARsolo \cite{kaminow2021starsolo} configured against the specific barcode-UMI read structure of the assay. Mapping to genes was configured in a forward-strand-specific manner and using the Gene assignment mode (as opposed to GeneFull), thus accounting for exonic overlaps only, as this is the model that is most compatible with the 3’ nature of the assay and has empirically proved to retrieve the most reads for quantification. Raw counts were normalised using sample-specific size factors adapted from the median-of-ratios method used in DESeq2 \cite{love2014differential}.}

\subsubsection{Proteomics}
Cells in 384-well plates were washed 3 times with PBS using a Bravo automated liquid handler (Agilent Technologies) to remove the residual culture media. Subsequently, 90 µL of 80\% acetone were added to the cell pellets and the plates were incubated at -20 C for 2h to precipitate the proteins. After centrifugation at 6000 g for 15 min, supernatants were removed and protein pellets were resuspended in 20 µL digestion buffer (100 mM TEAB (pH 8.5) containing 0.625 mM TCEP, 2.5 mM chloroacetamide, 3.125 ng/µL trypsin, and 3.125 ng/µL LysC) followed by over-night incubation at room temperature in an orbital shaker set to 1200 rpm. Peptides were dried in vacuo and labeled with TMTpro isobaric mass tags in 100 mM TEAB (pH 8.5) 50\% DMSO at room temperature for 1h. The reaction was stopped in 2.5\% hydroxylamine and samples were subsequently pooled and purified using C18SCX stage-tips as described \corrected{previously} \cite{werner2021affinity}. TMT-labeled samples were subjected \corrected{to} IMAC enrichment using prefilled Fe(III) cartridges (Agilent) and the flow-through was further processed with a high pH reversed-phase peptide fractionation kit (Thermo Fisher Scientific) to yield 3 fractions prior to LC-MS/MS analysis. 

\subsubsection{LC-MS/MS analysis of proteomics samples}
Fractionated and lyophilized samples were resuspended in 0.05\% trifluoroacetic acid in water and 30\% of each sample was injected into an Ultimate3000 nanoRLSC (Dionex) coupled to a Exploris (Thermo Fisher Scientific) mass spectrometer (Thermo Fisher Scientific). Peptides were separated on custom- made 50 cm × 100 µm (ID) reversed-phase columns (C18, 1.9 µm, Reprosil-Pur, Dr. Maisch) at 55 C. Gradient elution was performed from 2\% acetonitrile to 40
\% acetonitrile in 0.1\% formic acid and 3.5\% DMSO over 65 min at a flow rate of 350 nL/min. Samples were online injected into the mass spectrometer. The Q Exactive Plus was operated in a data-dependent top 10 acquisition method. MS spectra were acquired using 70.000 resolution and an ion target of 3 x 106 for MS1 scans. Higher energy collisional dissociation (HCD) scans were performed with 35\% NCE at 35.000 resolution (at m/z 200), and ion target setting was set to 2 x 105 to avoid coalescence \cite{werner2021affinity}. The instrument was operated with Tune 2.4 and Xcalibur 3.0 build 63. Phosphoproteomics samples were run unfractionated using a 120 min gradient and otherwise identical settings.

\subsubsection{Protein Identification and quantification}
Raw data were processed using an in-house pipeline based on the isobar quant package \cite{franken2015thermal}. Mascot 2.5 (Matrix Science, Boston, MA) for protein identification. In a first search 30 ppm peptide precursor mass and 30 mDa (HCD) mass tolerance for fragment ions was used for recalibration followed by search using a 10 ppm mass tolerance for peptide precursors and 20 mDa (HCD) mass tolerance for fragment ions. Enzyme specificity was set to trypsin with up to three missed cleavages. The search database consisted of the SwissProt sequence database (SwissProt Human release December 2018, 42 423 sequences) combined with a decoy version of this database created using scripts supplied by Matrix Science. Carbamidomethylation of cysteine residues and TMT modification of lysine residues were set as fixed modification. Methionine oxidation, and N- terminal acetylation of proteins, and TMT modification of peptide N-termini were set as variable modifications. For phosphoproteomics samples, serine, threonine, and tyrosine phosphorylation was set as variable modification. Unless stated otherwise, we accepted protein identifications as follows. (i) For single-spectrum to sequence assignments, we required this assignment to be the best match and a minimum Mascot score of 30 and a 10× difference of this assignment over the next best assignment. Based on these criteria, the decoy search results indicated \textless{}1\% false discovery rate (FDR). (ii) For multiple spectrum-to-sequence assignments and using the same parameters, the decoy search results indicate
\textless{}0.1\% FDR. All identified proteins were quantified; FDR for quantified proteins was below 1\%.

Reporter ion intensities were read from raw data and multiplied with ion accumulation times (in milliseconds) to yield a measure proportional to the number of ions; this measure is referred to as ion area. Spectra matching to peptides were filtered according to the following criteria: mascot ion score \textgreater{}15, signal-to-background of the precursor ion \textgreater{}4, and signal-to-interference \textgreater{}0.5 \cite{savitski2013measuring}. Fold-changes were corrected for isotope purity as described and adjusted for interference caused by co-eluting nearly isobaric peaks as estimated by the signal-to-interference measure \cite{savitski2010targeted}. Protein quantification was derived from individual spectra matching to distinct peptides by using a sum-based bootstrap algorithm; 95\% confidence intervals were calculated for all protein fold-changes that were quantified with more than three spectra.

\begin{figure}[t]
    \centering
    \includegraphics[width=1.0\textwidth]{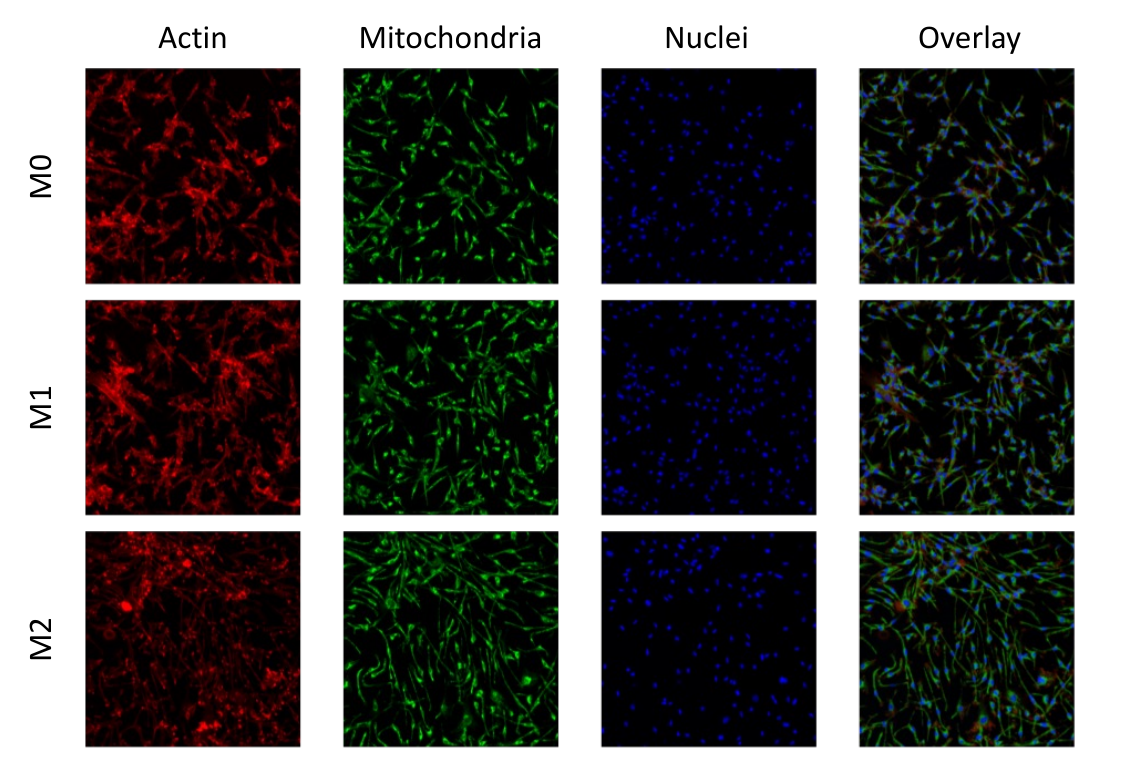}
    \caption{\textbf{Fluorescent dyes}. Typical confocal fluorescence images of actin \corrected{(phalloidin)} mitochondria \corrected{(MitoTracker)}, nuclei \corrected{(Hoechst)} and their overlay from left to right, respectively. Each row shows a different stimulation of the non-targeted (NT) control gRNA for visual comparison.}
    \label{fig:imaging_samples}
\end{figure}

\subsection{Automated High-Content Imaging}
\label{sec:automated_high_content_imaging}
High-content imaging (HCI) facilitates screening multiple cells at subcellular resolution to detect intensity, texture and morphology related phenotypic variation \cite{bray2016cell}. A PerkinElmer Opera Phenix high content screening system coupled with a collaborative robot designed for pharmaceutical screening (the plate::handler\uplett{TM} FLEX, PerkinElmer) was used to image \corrected{eight} 384-well plates \corrected{(four M1-stimulated and four M2-stimulated) in total} in spinning disk confocal mode. A 20x water immersion objective (NA=1.0) was used to collect 16 fields of view from each well \corrected{using excitation and emission filters appropriate for each dye, including Hoechst (for nuclei; excitation: 405 nm, emission: 435 nm-480 nm), MitoTracker (mitochondria; excitation: 561 nm, emission: 570 nm-630 nm), and phalloidin (actin, excitation: 640 nm, emission: 650 nm-760 nm).}

Two sCMOS cameras of the system were employed with pixel binning of 2. Each camera acquired a set of 2 fluorescence channels at a time (405 nm and 561 nm, 488 nm and 640 nm) and on a third acquisition, one of the cameras collected \corrected{a} brightfield (BF) images in transmission mode. Plates were imaged un-sealed but \corrected{with a} transparent plastic lid on, enabling BF transmission. A set of representative images from each channel of aforementioned phenotypic stain panel can be seen in \Cref{fig:imaging_samples}. All fluorescence imaging data was collected, visualized, and analysed for quality control using Harmony (PerkinElmer version 4.9.2137.273, Revision: 147881, Acapella version: 5.0.1.124082) software installed on the imaging instrument computer. Raw image files were annotated within the software defining plate maps with gene edits, donor codes, \corrected{and stimuli.}

\corrected{\paragraph*{Dye selection.} It should be noted that, while the experiments presented in this study employed three phenotypic dyes (Hoechst for nuclei, phalloidin for the actin cytoskeleton, and MitoTracker for the mitochondria) for high content image data generation, a typical phenotypic labeling protocol is likely to include up to 5 dyes (occasionally more, though spectral overlap and limitations of filter sets in most high content imagers traditionally cap the multiplexability at 5 fluorescent channels). The inclusion of additional dyes may enable improved model performance by adding to the granularity of the phenotype defined by high content imaging. Moreso, while a standard set of dyes like those from a Cell Painting approach \cite{bray2016cell} would facilitate model generalizability, tailoring the labeling panel could enable tuned model training for better predictability of specific pathways or cell types and under disease-relevant stimuli.}

\corrected{\paragraph*{Data preprocessing.} Low-expressed transcripts and proteins were filtered out by setting a threshold on the normalized counts. If the normalized counts were greater than zero in less than 50 out of 384 wells (13 \%) in each plate, the transcript/protein was eliminated from the analysis.}

\begin{figure}
    \centering\vspace{-2.5em}
    \begin{subfigure}[b]{\textwidth}
        \centering
        Transcriptomics (M1)
    \end{subfigure}
        \includegraphics[width=0.63\textwidth]{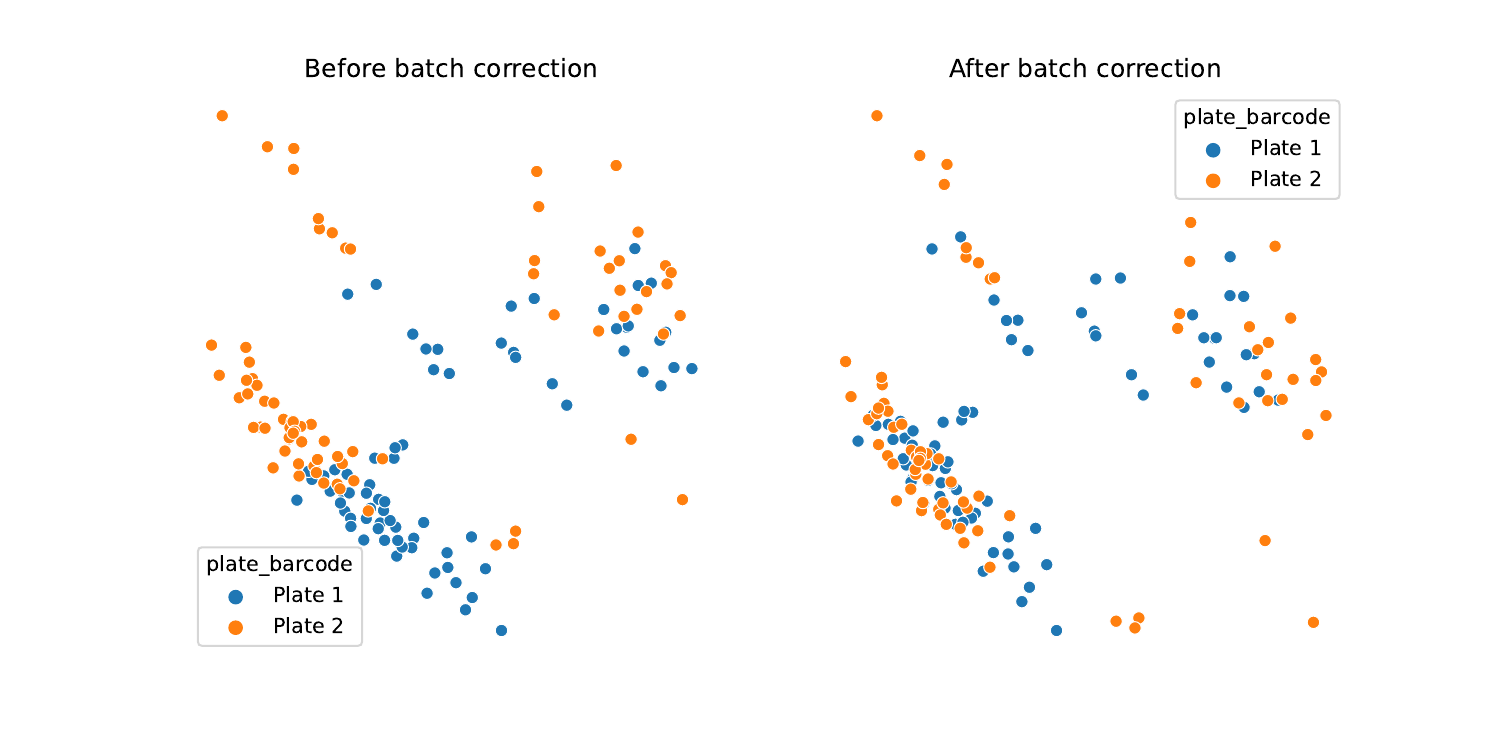}
        
    \begin{subfigure}[b]{\textwidth}
        \centering
        Transcriptomics (M2)
    \end{subfigure}
        \includegraphics[width=0.63\textwidth]{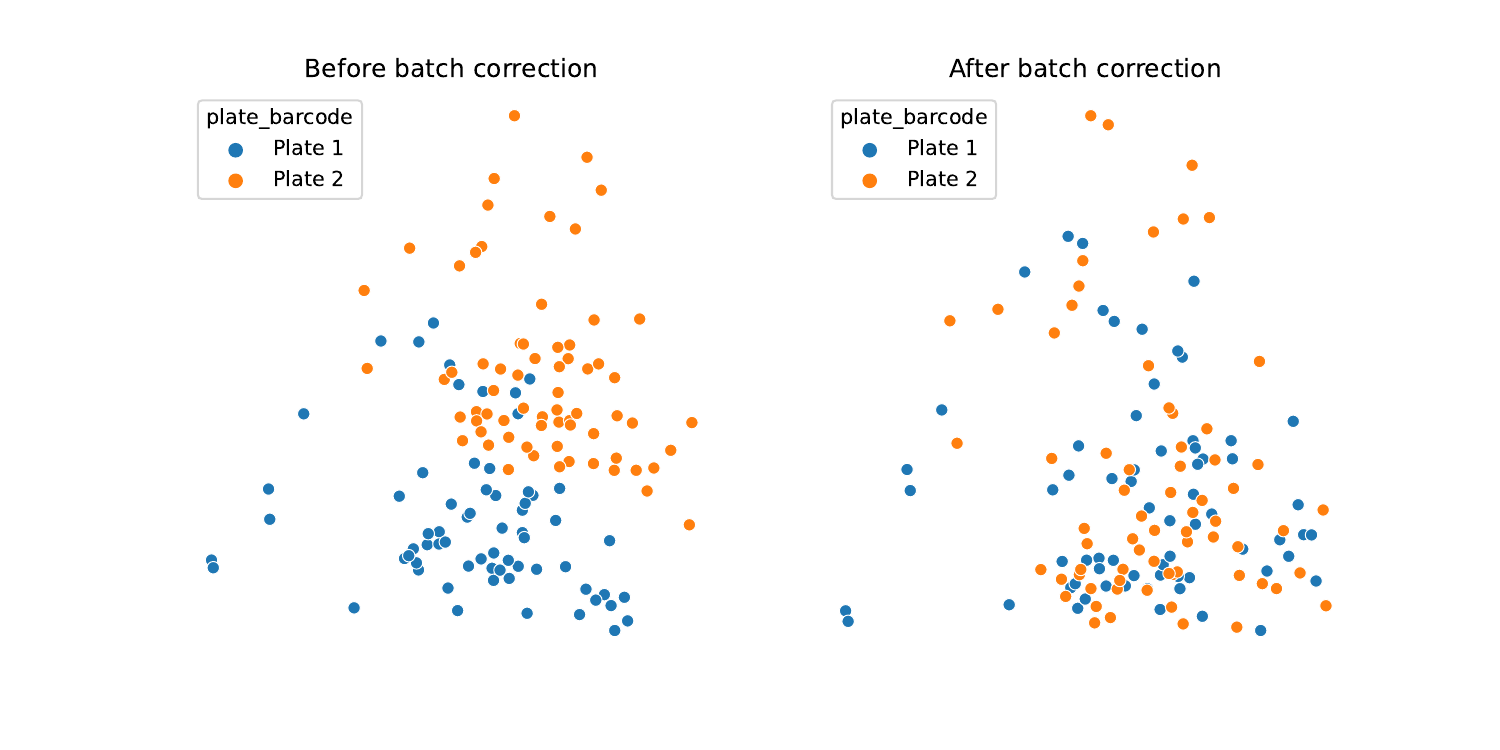}
        
    \begin{subfigure}[b]{\textwidth}
        \centering
        Proteomics (M1)
    \end{subfigure}
        \includegraphics[width=0.63\textwidth]{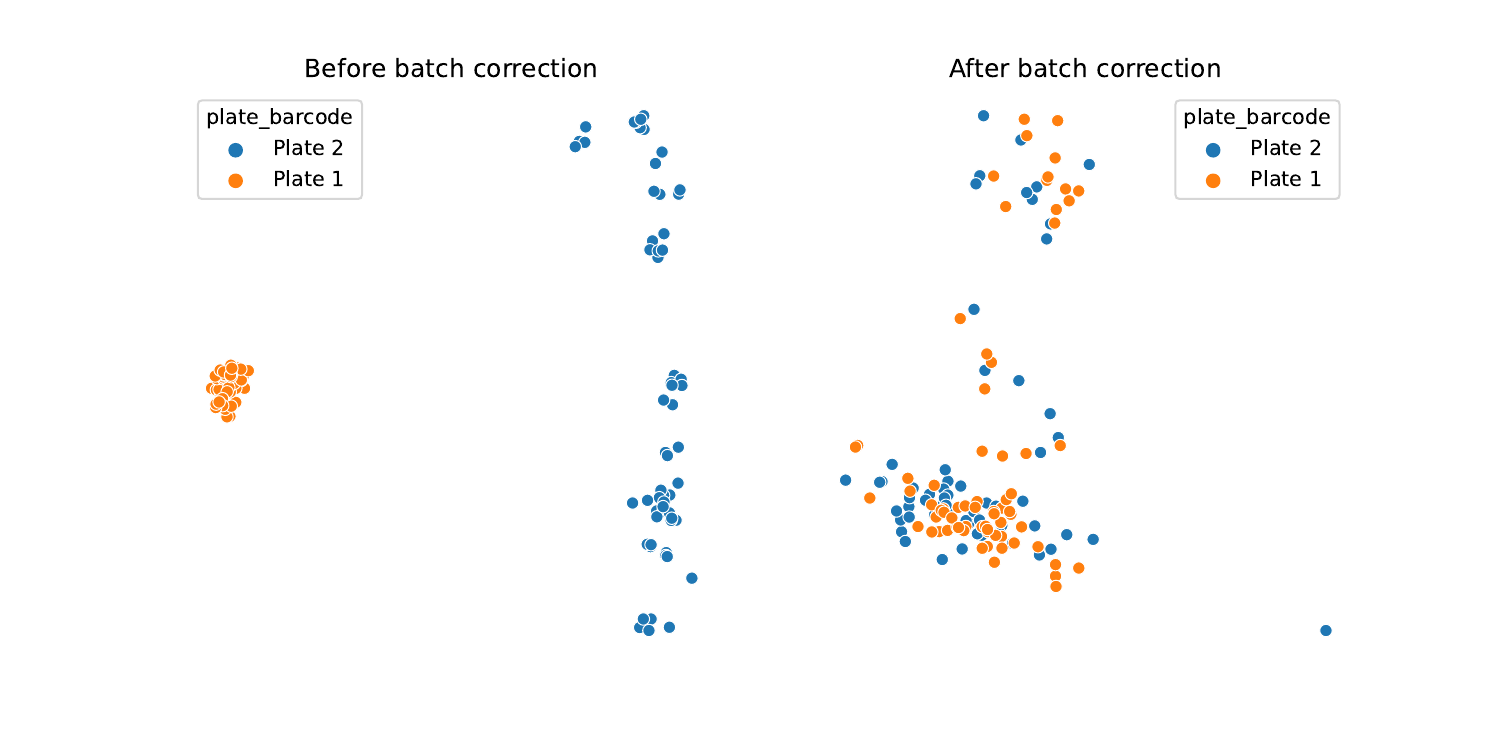}
        
    \begin{subfigure}[b]{\textwidth}
        \centering
        Proteomics (M2)
    \end{subfigure}
        \includegraphics[width=0.63\textwidth]{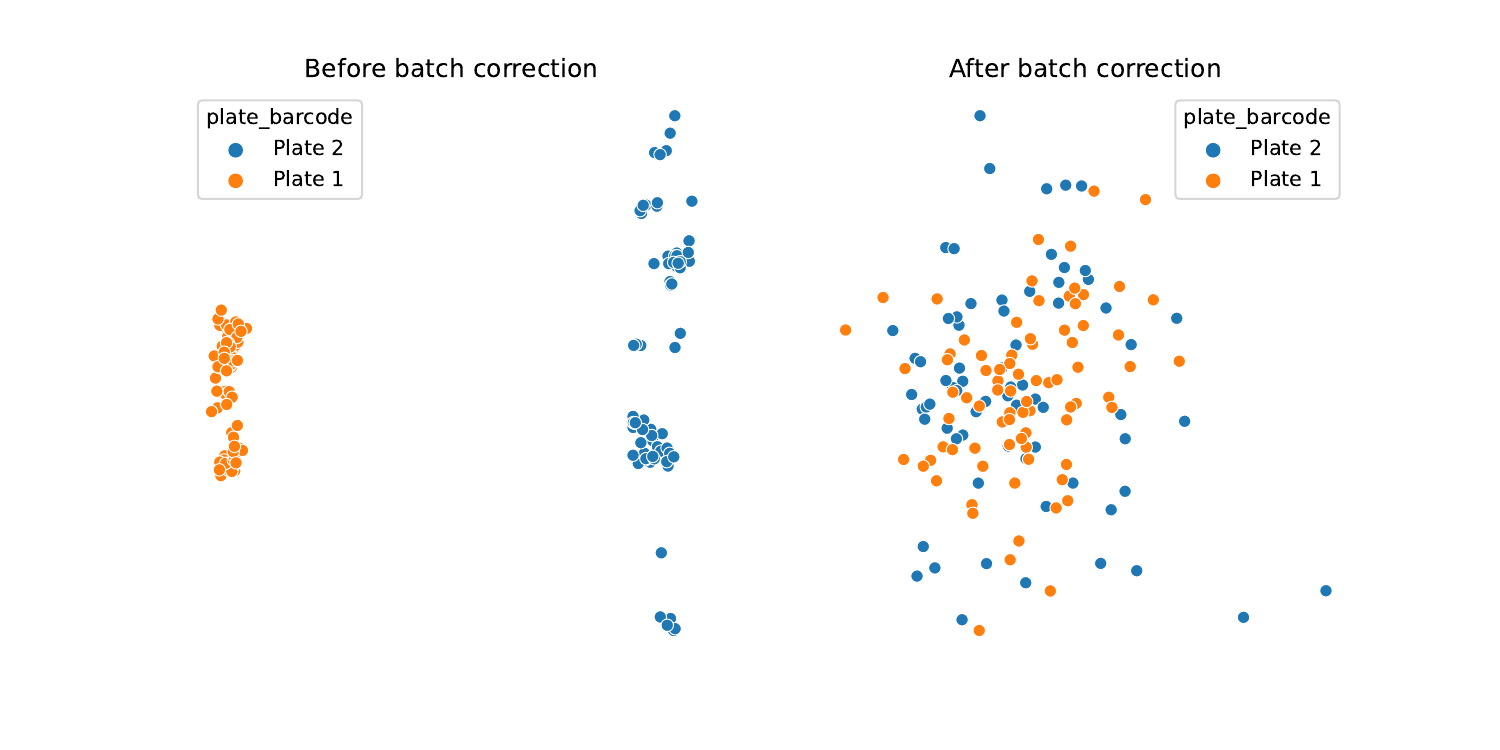}\vspace{-2em}
       \caption{\textbf{Qualitative assessment of plate-level batch effects.} Assessment of potential batch effects in transcriptomics and proteomics for M1 stimulated and M2 stimulated plates. Dots correspond to a principal component analysis (PCA) embedding of control wells in each of the two plates used for each stimulation condition and omics modality (orange and blue). Before batch correction (left hand plots for each omics modality and stimulation condition), batch effects are observable in control wells between plates of the same stimulation condition and omics type. Batch effects were removed (right hand for each omics modality and stimulation condition) after applying batch correction steps.
       }
       \label{fig:platebatch}
\end{figure}

\correctednn{\paragraph*{Normalization and batch correction of proteomics and transcriptomics.}
TMT-based proteomics and RNAseq transcriptomics data were normalized and batch corrected separately for M1 and M2 stimulated plates.  
For proteomics data, quantifications were normalized using variance stabilization normalization (R-package \texttt{vsn}) \cite{huber2002variance}. Subsequently, batch effects originating from TMT-multiplexing and differences between plates were corrected using the function \texttt{removeBatchEffect()} from R-package \texttt{limma} \cite{ritchie2015limma}.
For transcriptomics data, quantifications were normalized using sample-specific size factors adapted from the median-of-ratios method used in the R package \texttt{DESeq2} \cite{love2014moderated}. Subsequently, batch correction was performed as for proteomics data by correction for differences between plates using the function \texttt{removeBatchEffect()} from the R package \texttt{limma} applied to log-transformed normalized counts. Using the control wells on each plate, we verified qualitatively that plate batch effects were removed after batch correction and normalization (\Cref{fig:platebatch}).}

\subsection{Image2Omics}
\label{sec:imageanalysis}

\begin{figure}[t!]
    \centering
    \includegraphics[width=\linewidth]{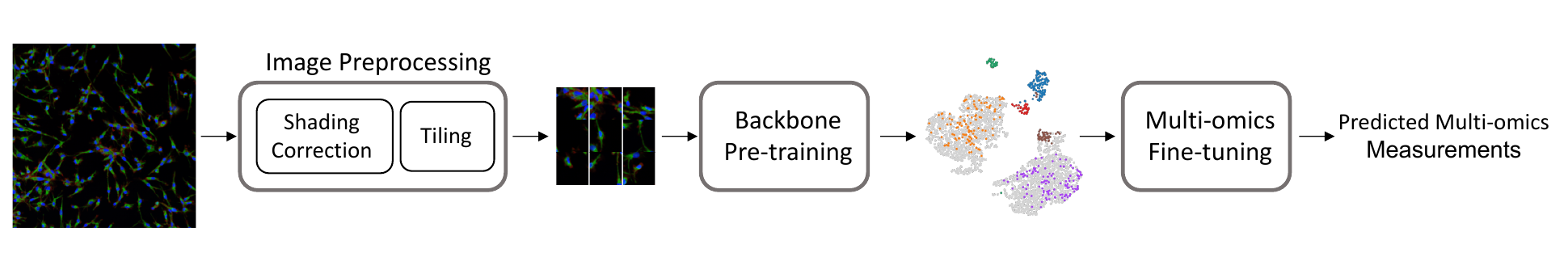}
    \caption{\textbf{Overview of the \themethod{} machine learning pipeline.} \corrected{Raw high content images (left) are preprocessed and tiled, fed into a backbone model for pretraining that is then fine-tuned to produce the final output predictions for transcript and protein abundances (right).}}
    \label{fig:pipeline}
\end{figure}

The Image2Omics pipeline, shown in \Cref{fig:pipeline}, consists of three stages: image preprocessing, backbone pre-training, and fine-tuning for multi-omics prediction. We describe each stage in detail in the following sections.

\subsubsection{Image Preprocessing}
\begin{figure}[t!]
    \centering
        \includegraphics[width=0.31\textwidth]{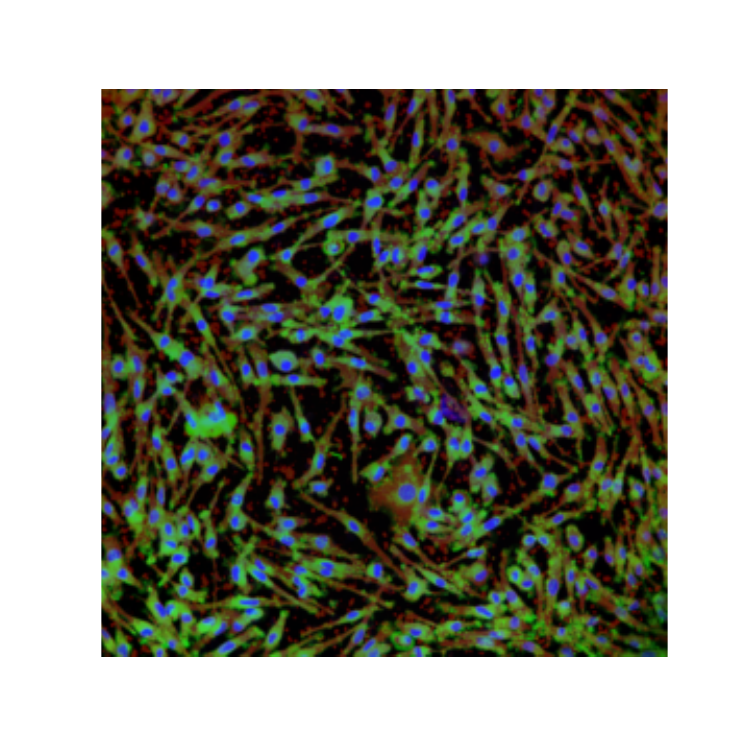}\quad
        \includegraphics[width=0.31\textwidth]{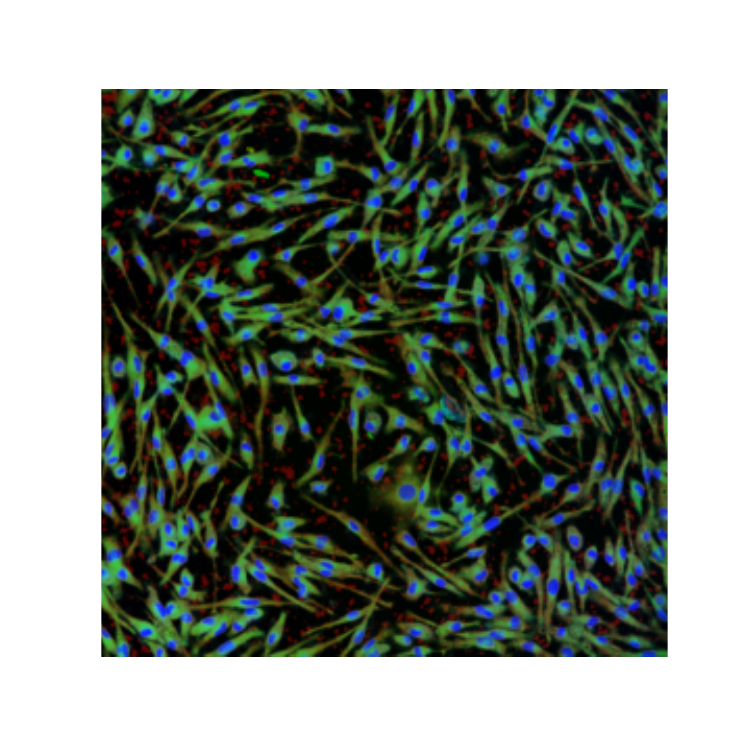}\quad
        \includegraphics[width=0.31\textwidth]{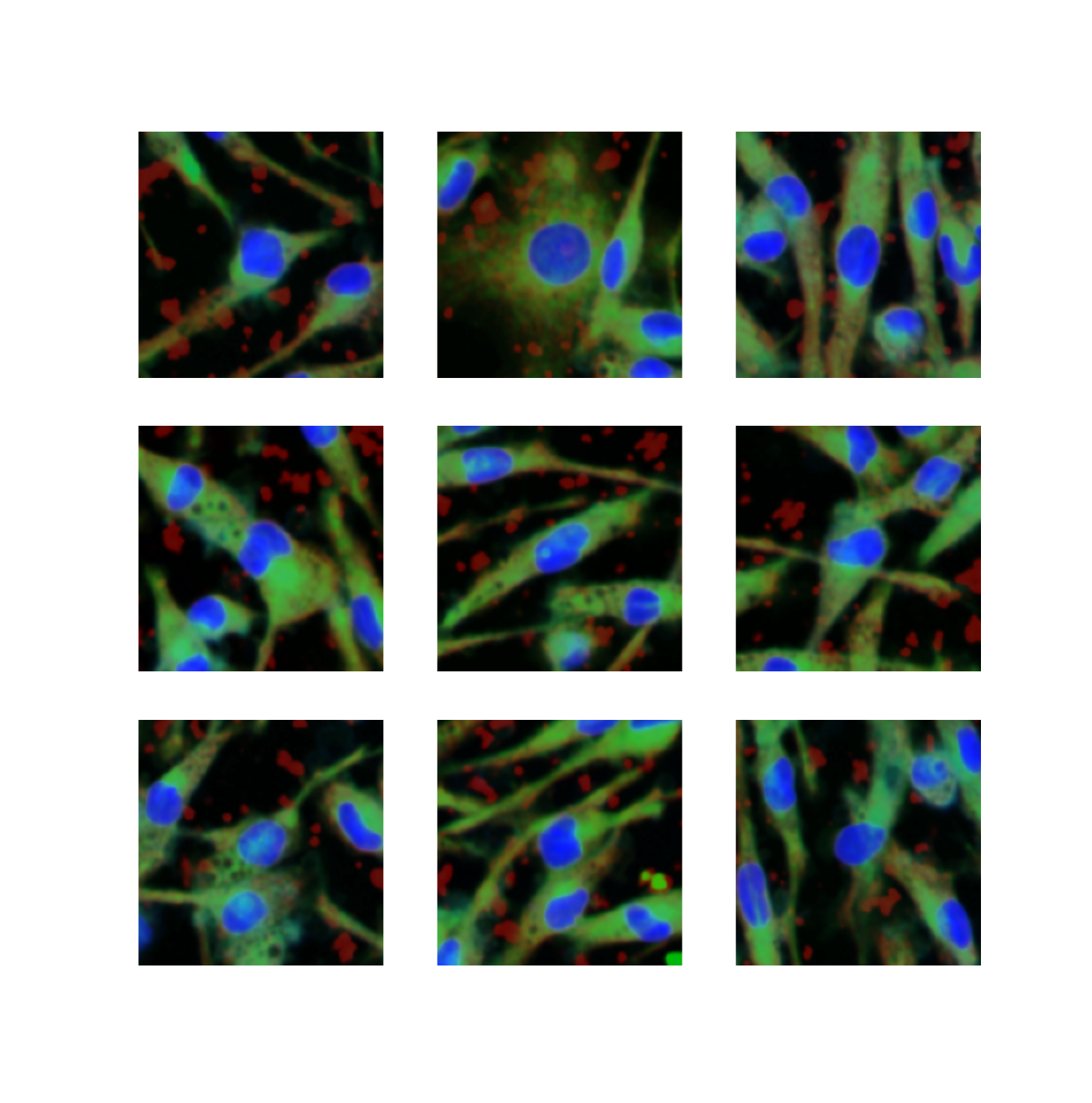}
       \caption{\textbf{Image preprocessing.} Acquired images (left) are first preprocessed by shading correction (center) and then separated into cell-centered patches (right).}
       \label{fig:preprocessing_before_after}
\end{figure}

\begin{figure}[h]
    \centering
           \begin{subfigure}[b]{0.05\textwidth}
        \centering
        \raisebox{0.0em}{\hspace{2.2em}}
    \end{subfigure}
    \begin{subfigure}[b]{0.4\textwidth}
        \raisebox{0.0em}{\hspace{1.7em}Most predictable}
    \end{subfigure}
    \begin{subfigure}[b]{0.3\textwidth}
        \raisebox{0.0em}{\hspace{-0.75em}Least predictable}
    \end{subfigure}
    
    \begin{subfigure}[b]{0.03\textwidth}
        \centering
        \raisebox{4.5em}{\rotatebox[origin=t]{90}{\hspace{2em}Transcriptomics (M1)}}
    \end{subfigure}
      \includegraphics[width=0.35\linewidth]{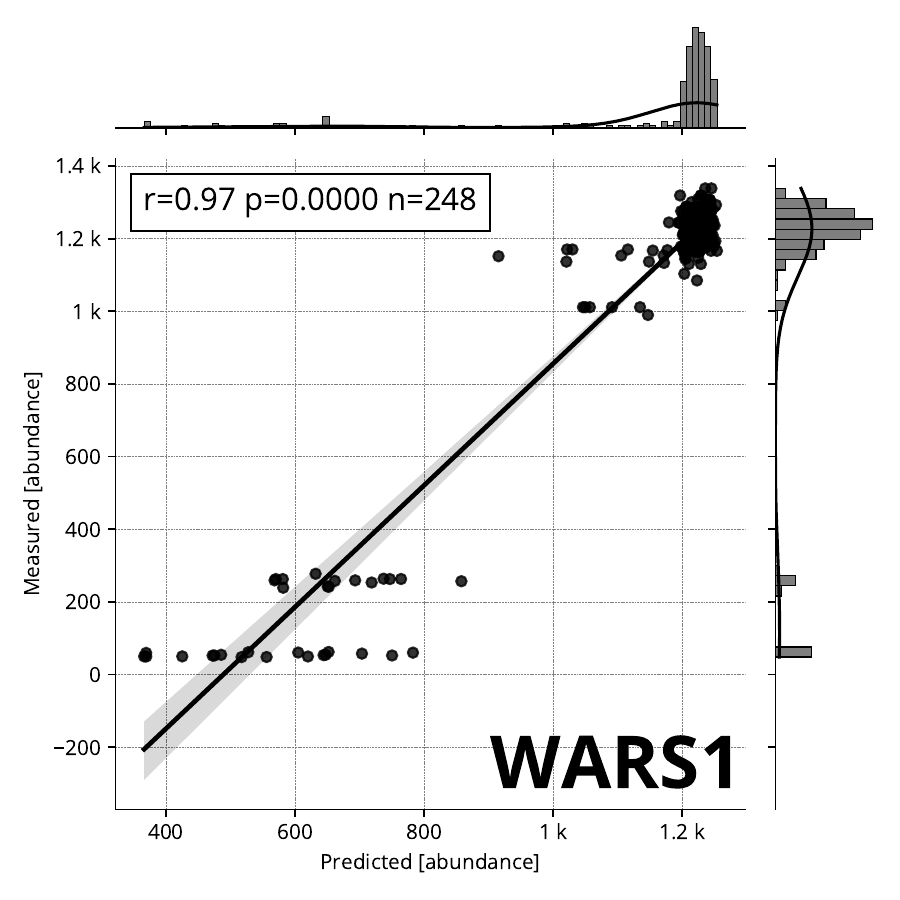}
      \includegraphics[width=0.35\linewidth]{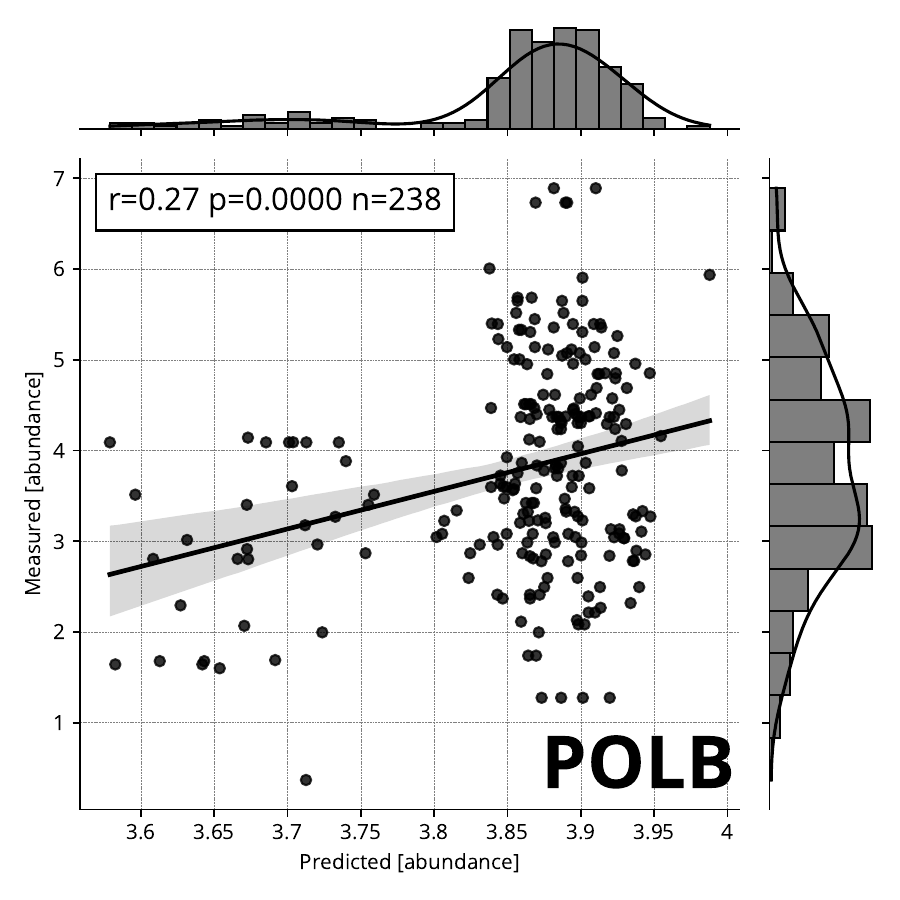}\quad
      
      \begin{subfigure}[b]{0.03\textwidth}
        \centering
        \raisebox{4.5em}{\rotatebox[origin=t]{90}{\hspace{2em}Transcriptomics (M2)}}
    \end{subfigure}
      \includegraphics[width=0.35\linewidth]{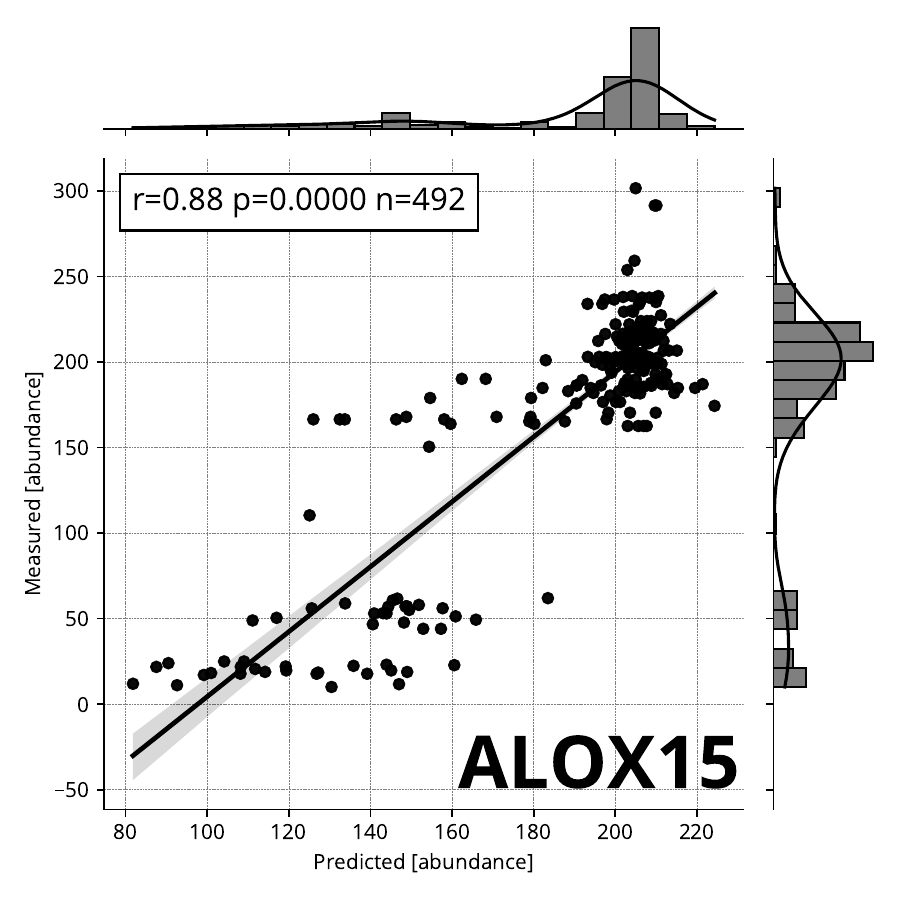}
      \includegraphics[width=0.35\linewidth]{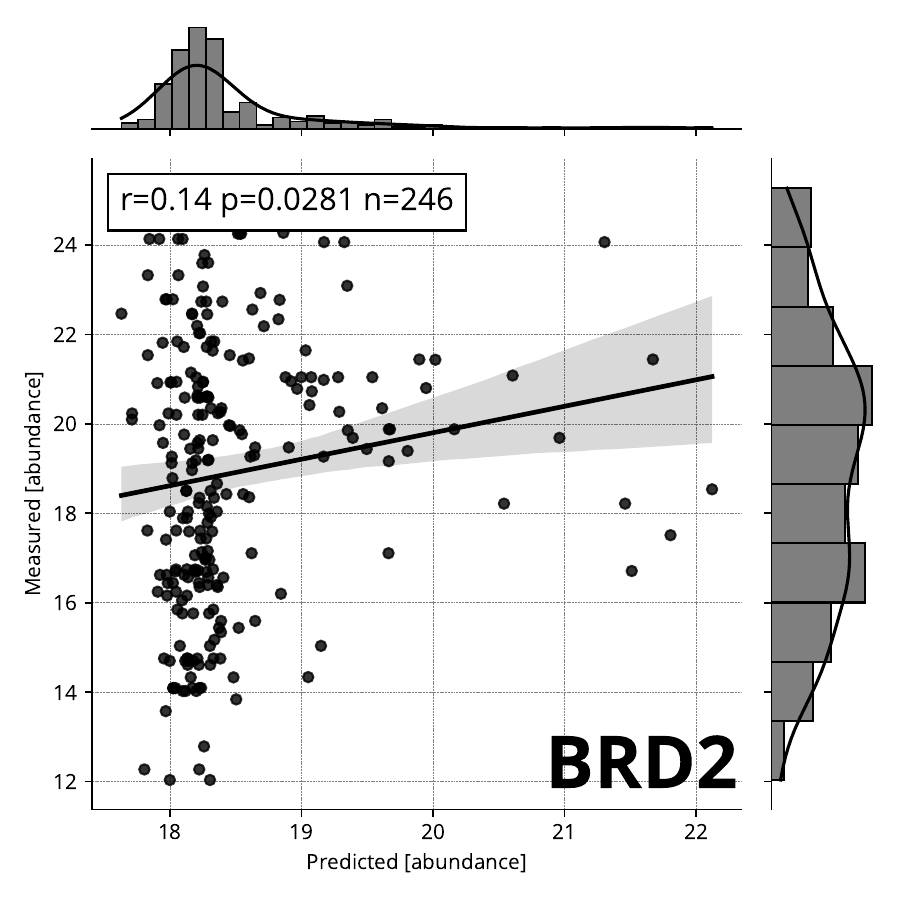}\quad
      
    \begin{subfigure}[b]{0.03\textwidth}
        \centering
        \raisebox{4.5em}{\rotatebox[origin=t]{90}{\hspace{2em}Proteomics (M1)}}
    \end{subfigure}
      \includegraphics[width=0.35\linewidth]{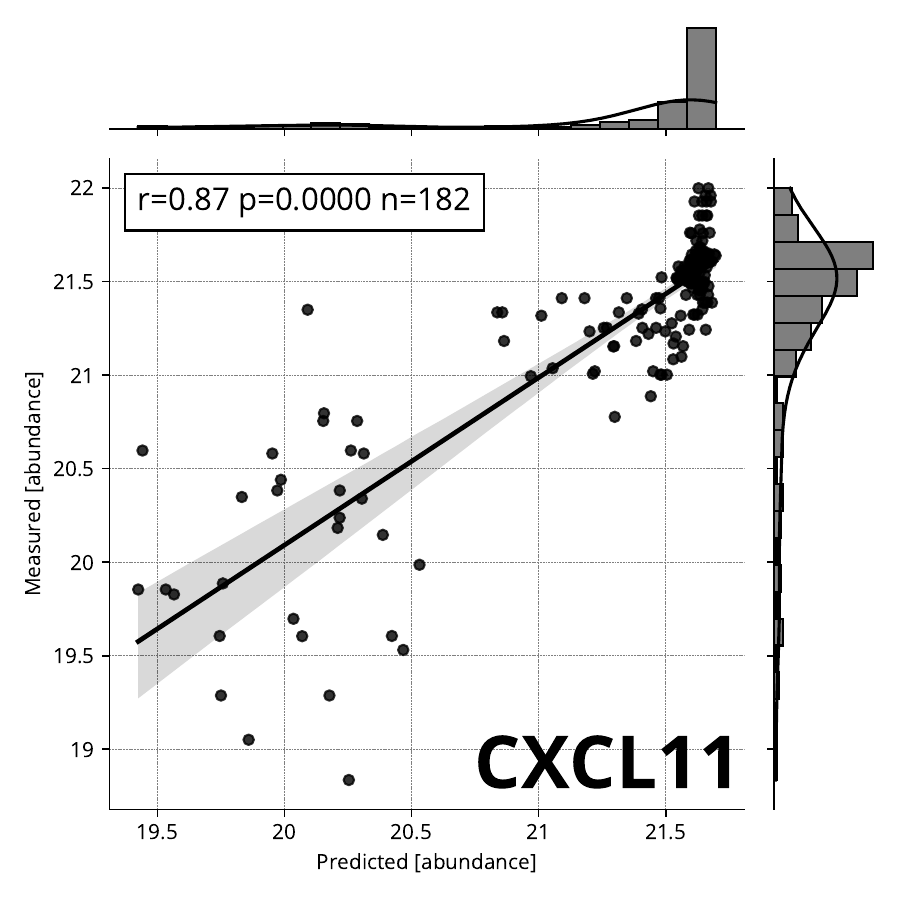}
      \includegraphics[width=0.35\linewidth]{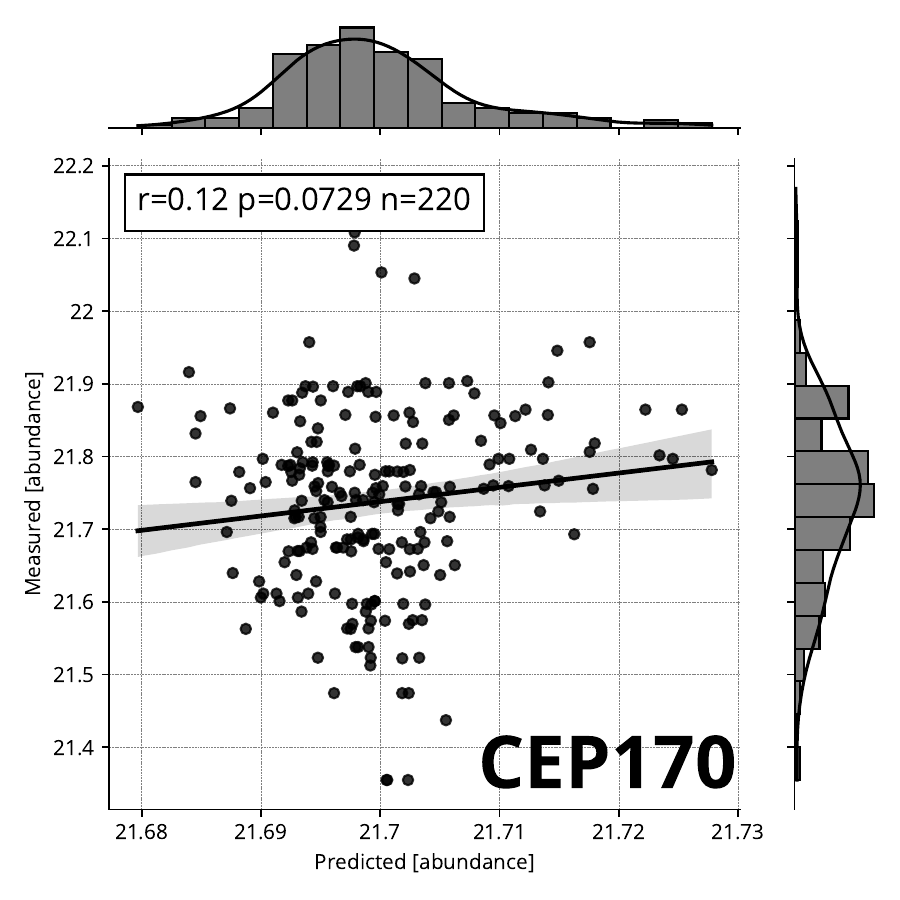}\quad
      
    \begin{subfigure}[b]{0.03\textwidth}
        \centering
        \raisebox{4.5em}{\rotatebox[origin=t]{90}{\hspace{2em}Proteomics (M2)}}
    \end{subfigure}
      \includegraphics[width=0.35\linewidth]{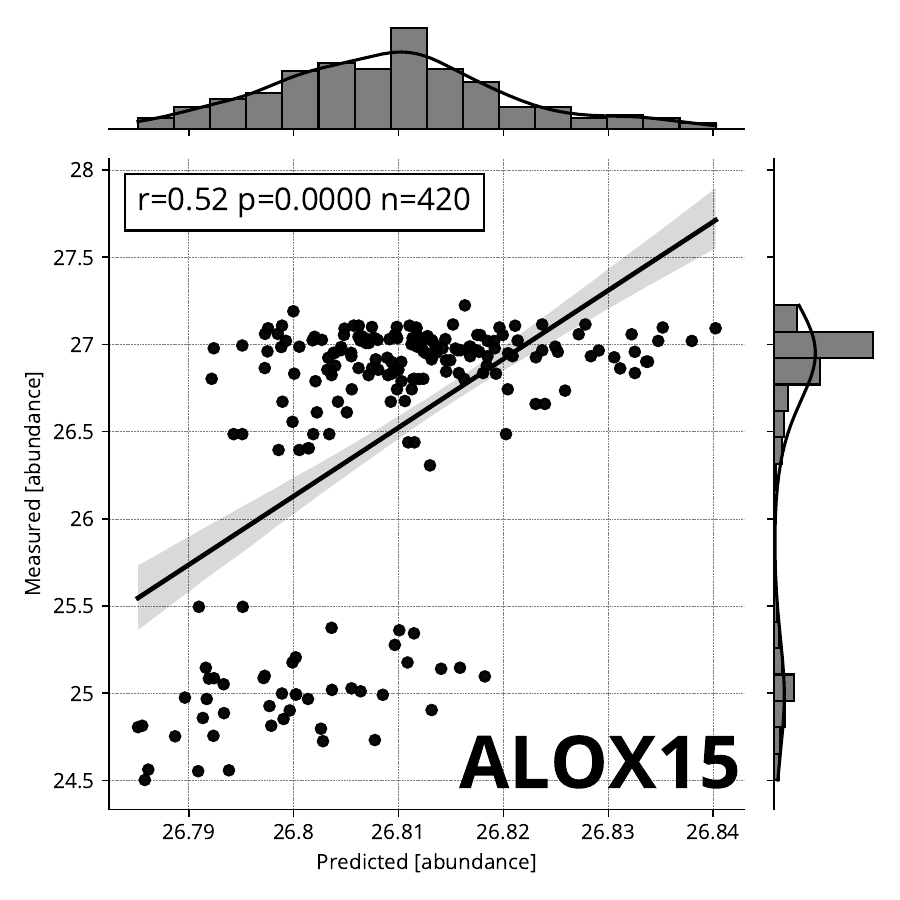}
      \includegraphics[width=0.35\linewidth]{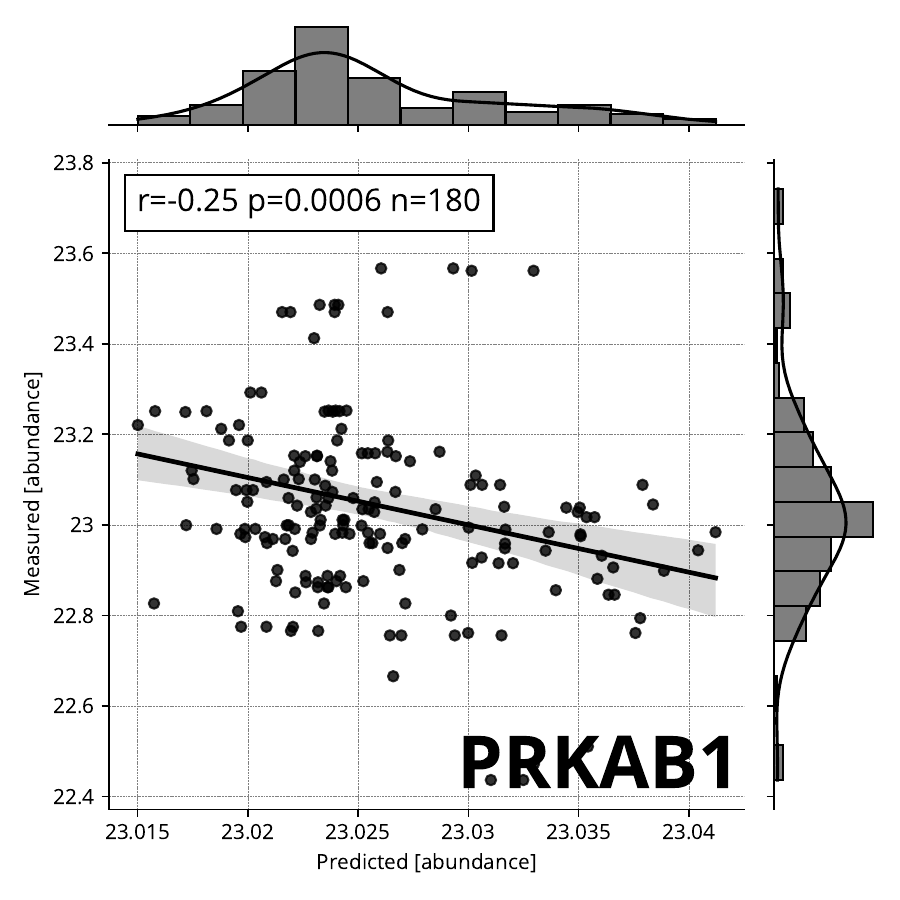}\quad
  
    \caption{\textbf{Most and least predictable gene products.} \corrected{The gene products from \Cref{tb:top_genes_performances_1} to \ref{tb:top_genes_performances_4} with the highest (left column) and lowest (right column) predictability and their respective associations between measured abundances and predicted abundances (each point represents a well in the dataset).}}
    \label{fig:scatter_plots_between_gene_products_and_predictability}
  \end{figure}

All acquired images are first preprocessed by (1) shading correction and (2) creating cell-centered patches (\Cref{fig:preprocessing_before_after}). In step (1), following \cite{singh2014pipeline}, shading correction for images in the $i$-$th$ plate is done according to
\[ I_{corrected}(x,y) = I_{measured}(x,y) \mathbin{/} F^i(x,y)\]
\[ F^i(x,y) = G_{\sigma} \ast \left(P_{k} \ast \dfrac{1}{N} \sum_{n=1}^{N} I_{measured,n}(x,y) \right)\]
where $x$ and $y$ denote pixel coordinates, $I_{corrected}$ and $I_{measured,n}$ are the corrected and measured intensity of the $n$-$th$ image out of $N$ total images in plate $i$-$th$, $F_p$ is the estimated flat-field image for plate  $i$-$th$, $G_{\sigma}$ is a Gaussian kernel with standard deviation $\sigma$, and $P_{k}$ is a $P$-percentile filter with kernel size $k$. Similar to \cite{ando2017improving}, we set $\sigma$ to 50 and use a $10th$-percentile filter with $k$ of 250.

In step (2), the cell-centered patches are created by identifying and cropping around nuclei centers from the corrected images, producing a $128\times128$ tile for each nuclei. To identify nuclei centers, we binarize the nucleus images using Otsu's method and apply Laplacian of Gaussian filter on the binarized image.

\subsubsection{Backbone Pre-training}
\label{sec:embedding_learning}
A deep convolutional neural network backbone is pre-trained to discriminate genetic perturbations from the processed images. Specifically, we use a $\texttt{ResNet18}$ \cite{he2016deep} backbone with three modifications: (i) we replace the {Average Pooling} and {Flatten} operations with a single {Global Average Pooling} layer to support arbitrary input image size, (ii) two dense layers of sizes 1024 and 128 are added following  {Global Average Pooling}, and (iii) we add an additional {Mean Aggregation} layer to enable multiple instance learning. Specifically, an embedding $h^k$ for a perturbation in well $k$ is calculated as
\[H_i^k = \texttt{mResNet18}(x_i^k) \]
\[H^k = \dfrac{1}{N} \sum_{i=1}^{N} H_i^k  \]
where $x_i^k \in \mathbb{R}^{128 \times 128 \times 3}$ and $H_i^k\in \mathbb{R}^{128}$ denote the $i$-$th$ cell-centered patch of well $k$ and its corresponding embedding, $\texttt{mResNet18}$ is the ResNet18 architecture with modifications (i) and (ii), and $H^k\in \mathbb{R}^{128}$ is the aggregated embedding from $N$ cell-centered patches in well $k$.

We train $mResNet18$ using Multiple Instance Triplet Loss, defined as
\[L = max(0, ||H -  H^+ ||^2_2 - || H -  H^- ||^2_2 + \alpha) \]
where $H$, $H^+$, and $H^-$ denote the aggregated embedding of the anchor, positive, and negative samples, respectively, and $\alpha$ is the margin term. Intuitively, the loss function optimizes for an embedding space where anchor and positive samples stay close to each other while anchor and negative samples are far apart.

Following \cite{schroff2015facenet}, we use semi-hard online triplet mining and define positive samples as wells that share same genetic perturbation and stimulation condition. During training, each batch consists of $K = 240$ wells uniformly distributed among 12 randomly sampled genetic perturbations and stimulation conditions. For each well we further randomly sample $N = 12$ instances for multiple instance learning, resulting in an effective batch size of 2880. We also set $\alpha = 0.2$ and use the AdamW optimizer with a learning rate of 0.0001 and weight decay term of 0.01.

\subsubsection{Multi-omics Fine-tuning}
\label{sec:predictor_network}
The pre-trained backbone is fine-tuned for multi-omics prediction - one model per omics modality. For each combination of omics modality $o \in \{\text{transcriptomics}, \text{proteomics}\}$ and stimulation  condition $s \in \{M1, M2\}$, we train a separate model with multiple instance learning following
\[Y^k_{o,s} = f_{\theta_{o,s}}\Biggl(\dfrac{1}{N} \sum_{i=1}^{N} \texttt{mResNet18}(x_i^k)\Biggl)\]
where $Y^k_{o,s}$ denote the predictions for well $k$, $f$ a linear model parametrized by $\theta_{o,s}$, and $x_i^k$ a cell-centered patch defined in \ref{sec:embedding_learning}. 

$\texttt{mResNet18}$ is initialized with pre-trained weights obtained in \ref{sec:embedding_learning} and remained frozen during training. For each batch, we sampled $K = 64$ wells and $N = 64$ patches from each well for an effective batch size of 4096. We again use the AdamW optimizer with a learning rate of 0.00005 and a weight decay term of 0.01 to minimize mean squared error loss between the predicted and actual gene product abundances.

\subsubsection{Inference}
The fine-tuned multi-omics models are employed for different omics modality and stimulation condition predictions. For each well, we sample N = 512 instances and predict the omics readouts using the fine-tuned models. 
Setting different random seeds can lead to different predictions due to selecting different set of instances. To make the results more robust, the prediction is repeated three times with different random seeds and the final results are the average across them. By choosing a big sample size (N = 512), we did not observe a significant variation across seeds.

\subsubsection{Potential integration of \themethod{} into experimental workflows}
{
A key requirement to adopt \themethod{} into experimental workflows is the availability of paired omics and cell imaging data from which to derive a base model that can be used to predict multi-omics measurements for a given cell population. In the absence of a pre-trained predictor, a typical experimental campaign may incorporate \themethod{} by initially collecting paired data until a sufficient training set is aggregated and subsequently substituting experimental measurements for the well-predictable omics modalities using \themethod{}. Under standardised imaging protocols, we envision that in the future sufficient data and pre-trained models may be available in the public domain for cell types of high interest to potentially immediately utilise \themethod{} without first having to collect paired omics and imaging data to derive a base model from. The exact amount of training data needed to achieve sufficiently accurate \themethod{} predictions is highly dependent on the downstream use case and the overall predictability of the desired set of markers from cell imaging. With use cases that interrogate the overall functional cell state having relatively lower data requirements compared to use cases that require the precise characterisation of the abundance of singular gene products. 
}

\subsection{Evaluation Protocol}
\label{sec:evaluation_protocol}
\subsubsection{Label Calculation}
Due to the many-to-many relationship between well replicates in imaging readout and well replicates in multi-omics readout, our label calculation protocol consists of three steps:
\begin{enumerate}
    \item For multi-omics readout, group the wells together based on the gene perturbation \corrected{and stimulation}.
    \item Randomly sample two wells per group and take the average of their labels.
    \item Use the average from the last step and assign it as a label representation for each group.
\end{enumerate}
Since in our dataset, there are two imaging readout plates per multi-omics readout plate, the above steps are repeated twice with two different random seeds to generate an imaging to multi-omics readout pair.
\subsubsection{Data splitting}
We evaluate Image2Omics by repeating the protocol in \ref{sec:imageanalysis} with 10 random splits. 
For data splitting, the split is done based on the gene perturbation \corrected{and stimulation condition}, where we randomly split the unique 156 gene perturbations into train/validation/test sets following 70/10/20 ratio. 
Due to using different random seeds for splitting, we will have a different set of wells in train/validation/test sets for each seed.
\subsubsection{Metrics}
We used Symmetric Mean Absolute Percentage Error (SMAPE) between the predicted and ground-truth gene product abundances over $K$ samples of the test set as the evaluation metric:

\[\text{SMAPE}(O, \hat{O}) = \frac{100\%}{K} \Sigma^{K}_{k=0} \frac{| \hat{O}_k - O_k |}{ |\hat{O}_k| + |O_k|} \]

We also performed statistical tests to evaluate the adequacy of the proposed model. For the statistical tests, we compared the predictions using our model with a static estimate $\bar{O}_i$ based on the training set, which was the average of the omics abundance value in the training set for the $i$-$th$ target. The comparison was done by applying Mann Whitney Wilcoxon test on the SMAPE obtained from our model versus the static training set estimate.  The null hypothesis for this is $\bar{x}i^1 = \bar{x}i^2$ with $\bar{x}i^1 = \text{SMAPE}(O, \hat{O})$ and $\bar{x}i^2 = \text{SMAPE}(O, \bar{O})$ where $K$ is the total number of images in the test set. The goal of this test was to determine the fraction of targets with significant predictive signal stemming from the cellular images. Rejecting the null hypothesis for a molecular marker would imply that \themethod{} does not have a lower SMAPE than a static training set estimate and that hence the molecular marker is not more predictable when having the cell imaging information available than in its absence.

\subsubsection{Quantitative comparison of \themethod{} features with cell morphology features}

\begin{figure}
    \vspace*{-18\baselineskip}
    \centering
    \includegraphics[width=0.975\textwidth]{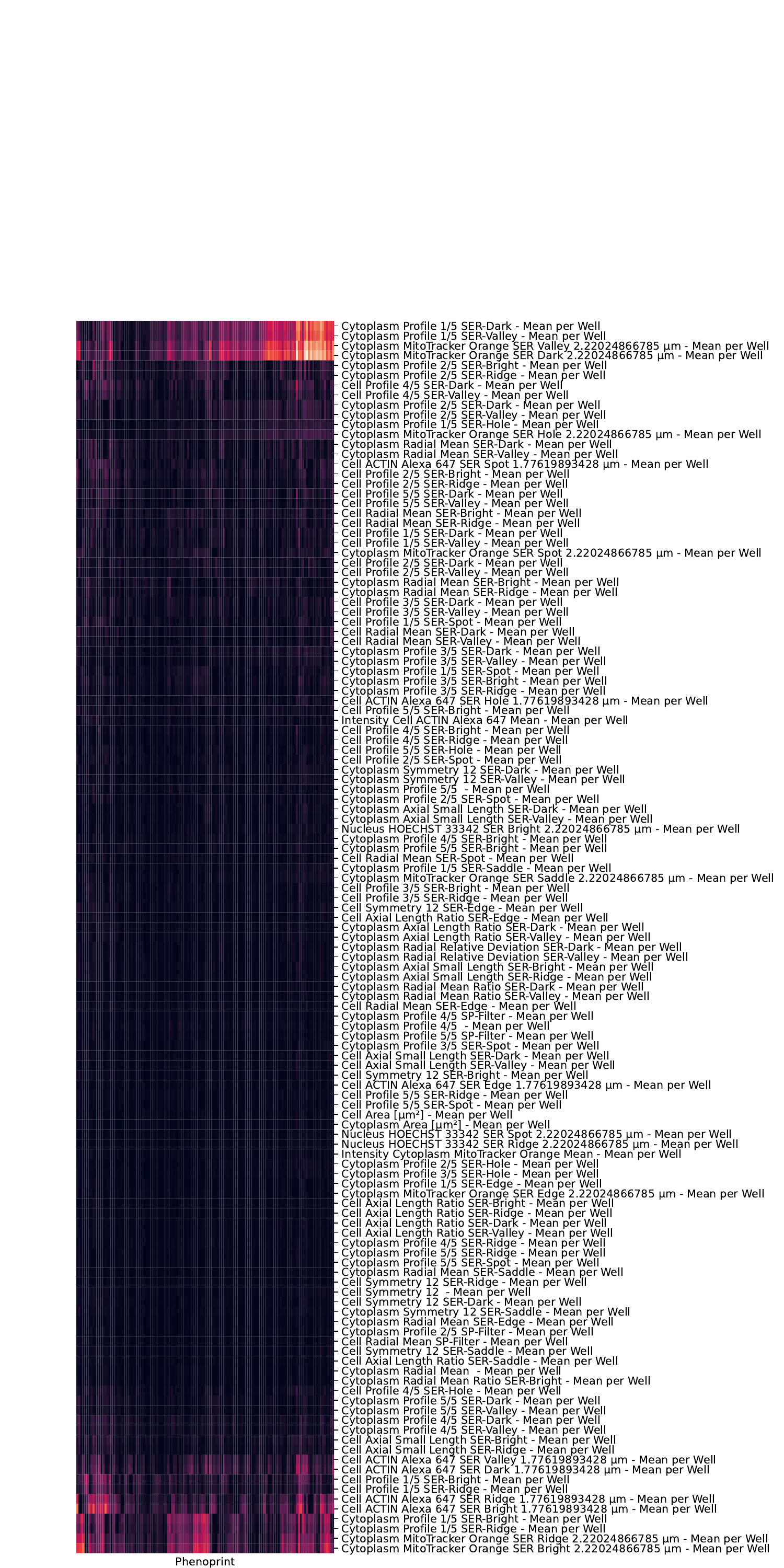}
   \caption{\textbf{Correlation with interpretable morphology features.} \corrected{Canonical correlation analysis (CCA) \cite{hardle2012canonical} of interpretable saddles, ridges and edges (SER) features generated by the Harmony software (PerkinElmer; y-axis) and their similarity (higher = brighter) with \themethod{} features (x-axis, indicated \enquote{Phenoprint}). Cytoplasm plasma membrane (profile 1/5) and MitoTracker orange under various intensity profiles (valley, ridge, bright and dark) are most correlated with \themethod{} features. We note that some learnt features may not correspond directly to interpretable features (i.e. may be novel).}}
    \label{fig:heatmap_harmony_features}
\end{figure}

\corrected{To aid in interpreting the feature embedding learnt by the image-embedding component of Image2Omics, we quantitatively investigated the correlation between features learnt by \themethod{} to interpretable cell morphology features generated by the Harmony (PerkinElmer, version 4.9.2137.273, Revision: 147881, Acapella version: 5.0.1.124082) software using Canonical Correlation Analysis (CCA) \cite{hardle2012canonical}. The results show that the highest similarity belongs to Cytoplasm profile, and Cytoplasm MitoTracker features, and the lowest similarity belongs to cell and cytoplasm symmetry and compactness, as well as number of round and longitudinal cells per well (\Cref{fig:heatmap_harmony_features}).}

\subsection{Data Availability}
\label{sec:data_availability}

Paired transcriptome, proteome and imaging data is made available in central repositories. Processed data can be accessed at s3://image2omics (see README at \url{https://github.com/GSK-AI/image2omics} for access instructions). Raw transcriptomics data are available on GEO GSE237485. Proteomics data are available via ProteomeXchange with identifier PXD044522.

\subsection{Code Availability}
\label{sec:code_availability}

The source code for \themethod{} is available at \url{https://github.com/GSK-AI/image2omics}.

\bibliography{main}
\bibliographystyle{naturemag}
\newpage

\begin{table}[h]
\centering
\vspace{-1.25ex}
\caption{Comparison of the top 10 and bottom 10 gene products with the respectively lowest and highest $r^2$ values (higher is better) and Symmetric Mean Absolute Percentage Errors (SMAPEs; lower is better, all significant at $\alpha = 0.05$) in predicting ground truth gene product abundances directly from cellular images in the test fold dataset for transcriptomics in M1 polarised macrophages. 95\% confidence intervals (CIs) were calculated via resampling and retraining models with a new random seed across $10$ resampled runs. Performance is calculated over all perturbed and unperturbed cell states to cover a diverse range of cellular states. }
\label{tb:top_genes_performances_1}
\vspace{1.0ex}
\begin{small}
\begin{tabular}{lrr}
\toprule
\multicolumn{3}{c}{Transcriptomics (M1 state)} \\
\midrule
\multicolumn{3}{c}{Top 10} \\
Gene symbol & $r^2$ & SMAPE (95\% CI) \\
\midrule
WARS1 & 0.91 (95\% CI: 0.83, 0.94) & 11.23\% (95\% CI: 9.89\%, 11.94\%) \\
SERPING1 & 0.90 (95\% CI: 0.70, 0.91) & 11.53\% (95\% CI: 10.09\%, 13.91\%) \\
ALOX5AP & 0.89 (95\% CI: 0.78, 0.94) & 12.38\% (95\% CI: 9.88\%, 14.10\%) \\
SLCO2B1 & 0.89 (95\% CI: 0.77, 0.90) & 14.88\% (95\% CI: 11.51\%, 18.84\%) \\
MAFB & 0.89 (95\% CI: 0.35, 0.91) & 12.08\% (95\% CI: 10.60\%, 15.58\%) \\
MMP2 & 0.89 (95\% CI: 0.76, 0.92) & 10.72\% (95\% CI: 9.54\%, 11.72\%) \\
IRF1 & 0.89 (95\% CI: 0.78, 0.91) & 11.01\% (95\% CI: 9.94\%, 11.91\%) \\
TAP1 & 0.88 (95\% CI: 0.83, 0.89) & 11.36\% (95\% CI: 10.20\%, 12.18\%) \\
PID1 & 0.88 (95\% CI: 0.81, 0.93) & 26.04\% (95\% CI: 25.05\%, 27.26\%) \\
C1orf162 & 0.88 (95\% CI: 0.79, 0.93) & 15.59\% (95\% CI: 14.30\%, 17.28\%) \\
\midrule
\multicolumn{3}{c}{Bottom 10} \\
Gene symbol & $r^2$ & SMAPE (95\% CI) \\
\midrule
POLB & 0.02 (95\% CI: 0.00, 0.16) & 13.29\% (95\% CI: 12.18\%, 14.09\%) \\
RPS15P4 & 0.02 (95\% CI: 0.01, 0.05) & 15.40\% (95\% CI: 13.67\%, 17.02\%) \\
PPP1R2 & 0.02 (95\% CI: 0.00, 0.07) & 7.93\% (95\% CI: 7.40\%, 8.79\%) \\
NAV1 & 0.02 (95\% CI: 0.00, 0.22) & 13.58\% (95\% CI: 11.53\%, 15.47\%) \\
TGFBR1 & 0.02 (95\% CI: 0.00, 0.06) & 22.74\% (95\% CI: 21.05\%, 25.91\%) \\
EMG1 & 0.02 (95\% CI: 0.00, 0.04) & 11.70\% (95\% CI: 10.30\%, 13.44\%) \\
GAS5 & 0.02 (95\% CI: 0.00, 0.03) & 17.55\% (95\% CI: 16.66\%, 19.26\%) \\
TPM1 & 0.02 (95\% CI: 0.00, 0.15) & 45.88\% (95\% CI: 41.46\%, 47.69\%) \\
STRN4 & 0.02 (95\% CI: 0.00, 0.04) & 8.84\% (95\% CI: 7.71\%, 9.69\%) \\
FBXW8 & 0.02 (95\% CI: 0.00, 0.31) & 12.68\% (95\% CI: 9.98\%, 14.72\%) \\
\bottomrule
\end{tabular}
\end{small}
\end{table}

\begin{table}[h]
\centering
\vspace{-1.25ex}
\caption{Comparison of the top 10 and bottom 10 gene products with the respectively lowest and highest $r^2$ values (higher is better) and Symmetric Mean Absolute Percentage Errors (SMAPEs; lower is better, all significant at $\alpha = 0.05$) in predicting ground truth gene product abundances directly from cellular images in the test fold dataset for transcriptomics in M2 polarised macrophages. 95\% confidence intervals (CIs) were calculated via resampling and retraining models with a new random seed across $10$ resampled runs. Performance is calculated over all perturbed and unperturbed cell states to cover a diverse range of cellular states. }
\label{tb:top_genes_performances_2}
\vspace{1.0ex}
\begin{small}
\begin{tabular}{lrr}
\toprule
\multicolumn{3}{c}{Transcriptomics (M2 state)} \\
\midrule
\multicolumn{3}{c}{Top 10} \\
Gene symbol & $r^2$ & SMAPE (95\% CI) \\
\midrule
ALOX15 & 0.77 (95\% CI: 0.64, 0.83) & 15.28\% (95\% CI: 14.26\%, 16.65\%) \\
APBB1IP & 0.77 (95\% CI: 0.66, 0.82) & 9.49\% (95\% CI: 8.89\%, 10.09\%) \\
VIM & 0.76 (95\% CI: 0.61, 0.84) & 5.26\% (95\% CI: 4.66\%, 5.90\%) \\
TGM2 & 0.75 (95\% CI: 0.61, 0.83) & 9.04\% (95\% CI: 8.35\%, 10.45\%) \\
C5AR1 & 0.75 (95\% CI: 0.65, 0.80) & 12.41\% (95\% CI: 11.19\%, 12.97\%) \\
PFKP & 0.74 (95\% CI: 0.62, 0.79) & 5.43\% (95\% CI: 5.18\%, 6.21\%) \\
ALOX5 & 0.74 (95\% CI: 0.59, 0.81) & 13.56\% (95\% CI: 12.06\%, 15.20\%) \\
CKB & 0.74 (95\% CI: 0.70, 0.80) & 11.04\% (95\% CI: 10.64\%, 12.44\%) \\
CYBB & 0.73 (95\% CI: 0.66, 0.78) & 8.66\% (95\% CI: 8.10\%, 9.20\%) \\
FCGR1A & 0.73 (95\% CI: 0.46, 0.80) & 9.70\% (95\% CI: 8.15\%, 12.07\%) \\
\midrule
\multicolumn{3}{c}{Bottom 10} \\
Gene symbol & $r^2$ & SMAPE (95\% CI) \\
\midrule
BRD2 & 0.02 (95\% CI: 0.00, 0.14) & 6.78\% (95\% CI: 5.69\%, 7.22\%) \\
SNRK & 0.02 (95\% CI: 0.00, 0.03) & 13.92\% (95\% CI: 12.06\%, 16.78\%) \\
SEPHS2 & 0.02 (95\% CI: 0.00, 0.05) & 5.86\% (95\% CI: 4.68\%, 6.99\%) \\
FOXRED1 & 0.02 (95\% CI: 0.00, 0.05) & 13.06\% (95\% CI: 12.09\%, 13.83\%) \\
SEC24A & 0.02 (95\% CI: 0.00, 0.06) & 10.28\% (95\% CI: 9.42\%, 10.68\%) \\
BMP1 & 0.02 (95\% CI: 0.00, 0.05) & 18.04\% (95\% CI: 15.83\%, 18.99\%) \\
PSMB1 & 0.02 (95\% CI: 0.00, 0.11) & 3.92\% (95\% CI: 3.18\%, 4.53\%) \\
HMCES & 0.02 (95\% CI: 0.01, 0.04) & 13.51\% (95\% CI: 11.60\%, 14.79\%) \\
DTX3L & 0.02 (95\% CI: 0.00, 0.08) & 11.67\% (95\% CI: 10.54\%, 13.61\%) \\
SMARCC2 & 0.02 (95\% CI: 0.00, 0.07) & 13.36\% (95\% CI: 11.93\%, 15.14\%) \\
\bottomrule
\end{tabular}
\end{small}
\end{table}

\begin{table}[h]
\centering
\vspace{-1.25ex}
\caption{Comparison of the top 10 and bottom 10 gene products with the respectively lowest and highest $r^2$ values (higher is better) and Symmetric Mean Absolute Percentage Errors (SMAPEs; lower is better, all significant at $\alpha = 0.05$) in predicting ground truth gene product abundances directly from cellular images in the test fold dataset for proteomics in M1 polarised macrophages. 95\% confidence intervals (CIs) were calculated via resampling and retraining models with a new random seed across $10$ resampled runs. Performance is calculated over all perturbed and unperturbed cell states to cover a diverse range of cellular states. }
\label{tb:top_genes_performances_3}
\vspace{1.0ex}
\begin{small}
\begin{tabular}{lrr}
\toprule
\multicolumn{3}{c}{Proteomics (M1 state)} \\
\midrule
\multicolumn{3}{c}{Top 10} \\
Gene symbol & $r^2$ & SMAPE (95\% CI) \\
\midrule
CXCL11 & 0.73 (95\% CI: 0.29, 0.75) & 0.60\% (95\% CI: 0.51\%, 0.99\%) \\
CXCL9 & 0.73 (95\% CI: 0.25, 0.74) & 0.41\% (95\% CI: 0.35\%, 0.69\%) \\
WARS & 0.72 (95\% CI: 0.20, 0.74) & 0.25\% (95\% CI: 0.24\%, 0.45\%) \\
CXCL10 & 0.71 (95\% CI: 0.16, 0.74) & 0.47\% (95\% CI: 0.41\%, 0.82\%) \\
GBP1 & 0.70 (95\% CI: 0.18, 0.74) & 0.35\% (95\% CI: 0.29\%, 0.58\%) \\
IDO1 & 0.69 (95\% CI: 0.16, 0.71) & 0.34\% (95\% CI: 0.29\%, 0.51\%) \\
CD38 & 0.68 (95\% CI: 0.24, 0.72) & 0.42\% (95\% CI: 0.35\%, 0.59\%) \\
GBP4 & 0.67 (95\% CI: 0.52, 0.69) & 0.27\% (95\% CI: 0.24\%, 0.43\%) \\
STAT1 & 0.67 (95\% CI: 0.43, 0.71) & 0.27\% (95\% CI: 0.25\%, 0.38\%) \\
GBP2 & 0.65 (95\% CI: 0.41, 0.68) & 0.28\% (95\% CI: 0.23\%, 0.44\%) \\
\midrule
\multicolumn{3}{c}{Bottom 10} \\
Gene symbol & $r^2$ & SMAPE (95\% CI) \\
\midrule
CEP170 & 0.03 (95\% CI: 0.01, 0.17) & 0.26\% (95\% CI: 0.24\%, 0.28\%) \\
RAB10 & 0.03 (95\% CI: 0.00, 0.09) & 0.37\% (95\% CI: 0.34\%, 0.52\%) \\
ALDOC & 0.03 (95\% CI: 0.00, 0.11) & 0.56\% (95\% CI: 0.51\%, 0.58\%) \\
AP2B1 & 0.03 (95\% CI: 0.00, 0.14) & 0.22\% (95\% CI: 0.18\%, 0.24\%) \\
MAP2K2 & 0.03 (95\% CI: 0.00, 0.16) & 0.24\% (95\% CI: 0.20\%, 0.31\%) \\
ANP32A & 0.03 (95\% CI: 0.01, 0.05) & 0.17\% (95\% CI: 0.16\%, 0.20\%) \\
C16orf62 & 0.03 (95\% CI: 0.00, 0.08) & 0.20\% (95\% CI: 0.19\%, 0.22\%) \\
OSBPL9 & 0.02 (95\% CI: 0.00, 0.11) & 0.18\% (95\% CI: 0.15\%, 0.19\%) \\
GPI & 0.02 (95\% CI: 0.00, 0.29) & 0.14\% (95\% CI: 0.12\%, 0.19\%) \\
PPIB & 0.02 (95\% CI: 0.00, 0.09) & 0.32\% (95\% CI: 0.26\%, 0.33\%) \\
\bottomrule
\end{tabular}
\end{small}
\end{table}

\begin{table}[h]
\centering
\vspace{-1.25ex}
\caption{Comparison of the top 10 and bottom 10 gene products with the respectively lowest and highest $r^2$ values (higher is better) and Symmetric Mean Absolute Percentage Errors (SMAPEs; lower is better, all significant at $\alpha = 0.05$) in predicting ground truth gene product abundances directly from cellular images in the test fold dataset for proteomics in M2 polarised macrophages. 95\% confidence intervals (CIs) were calculated via resampling and retraining models with a new random seed across $10$ resampled runs. Performance is calculated over all perturbed and unperturbed cell states to cover a diverse range of cellular states. }
\label{tb:top_genes_performances_4}
\vspace{1.0ex}
\begin{small}
\begin{tabular}{lrr}
\toprule
\multicolumn{3}{c}{Proteomics (M2 state)} \\
\midrule
\multicolumn{3}{c}{Top 10} \\
Gene symbol & $r^2$ & SMAPE (95\% CI) \\
\midrule
ALOX15 & 0.69 (95\% CI: 0.30, 0.82) & 0.67\% (95\% CI: 0.54\%, 1.11\%) \\
HSPH1 & 0.65 (95\% CI: 0.14, 0.81) & 0.14\% (95\% CI: 0.12\%, 0.24\%) \\
FABP4 & 0.64 (95\% CI: 0.52, 0.77) & 0.57\% (95\% CI: 0.47\%, 0.92\%) \\
EPB41L2 & 0.64 (95\% CI: 0.33, 0.73) & 0.25\% (95\% CI: 0.23\%, 0.37\%) \\
ANPEP & 0.63 (95\% CI: 0.29, 0.80) & 0.22\% (95\% CI: 0.19\%, 0.33\%) \\
TGM2 & 0.60 (95\% CI: 0.05, 0.70) & 0.31\% (95\% CI: 0.23\%, 0.51\%) \\
APBB1IP & 0.59 (95\% CI: 0.10, 0.71) & 0.45\% (95\% CI: 0.34\%, 0.63\%) \\
MAOA & 0.59 (95\% CI: 0.14, 0.71) & 0.36\% (95\% CI: 0.33\%, 0.54\%) \\
CKB & 0.59 (95\% CI: 0.14, 0.63) & 0.36\% (95\% CI: 0.32\%, 0.50\%) \\
CA2 & 0.54 (95\% CI: 0.39, 0.71) & 0.48\% (95\% CI: 0.42\%, 0.51\%) \\
\midrule
\multicolumn{3}{c}{Bottom 10} \\
Gene symbol & $r^2$ & SMAPE (95\% CI) \\
\midrule
PRKAB1 & 0.03 (95\% CI: 0.00, 0.06) & 0.32\% (95\% CI: 0.29\%, 0.38\%) \\
NNT & 0.03 (95\% CI: 0.00, 0.06) & 0.17\% (95\% CI: 0.15\%, 0.20\%) \\
PCBP1 & 0.02 (95\% CI: 0.00, 0.13) & 0.14\% (95\% CI: 0.11\%, 0.14\%) \\
RBMX & 0.02 (95\% CI: 0.00, 0.13) & 0.28\% (95\% CI: 0.23\%, 0.38\%) \\
FTL & 0.02 (95\% CI: 0.00, 0.04) & 0.40\% (95\% CI: 0.38\%, 0.44\%) \\
REEP5 & 0.02 (95\% CI: 0.00, 0.12) & 0.28\% (95\% CI: 0.23\%, 0.33\%) \\
HINT1 & 0.02 (95\% CI: 0.01, 0.10) & 0.16\% (95\% CI: 0.14\%, 0.19\%) \\
PSMC4 & 0.02 (95\% CI: 0.00, 0.03) & 0.12\% (95\% CI: 0.10\%, 0.17\%) \\
TPI1 & 0.02 (95\% CI: 0.00, 0.08) & 0.35\% (95\% CI: 0.32\%, 0.41\%) \\
DDX3X & 0.01 (95\% CI: 0.00, 0.13) & 0.18\% (95\% CI: 0.15\%, 0.22\%) \\
\bottomrule
\end{tabular}
\end{small}
\end{table}

\end{document}